\documentclass[twocolumn ]{aastex62}

\shorttitle{}
\shortauthors{Chouliaras et al.}

\usepackage{graphicx}
\usepackage{times}
\usepackage[]{natbib}
\usepackage{url}
\usepackage{color}
\usepackage{amssymb,amsmath}
\usepackage{xfrac}
\usepackage{float}
\usepackage{placeins}

\begin{document}

\title{Magnetic flux emergence and solar eruptions in partially ionized plasmas.}

\correspondingauthor{Georgios Chouliaras}
\email{gc205@st-andrews.ac.uk}

\author{Georgios Chouliaras}
\affil{School of Mathematics and Statistics, St. Andrews University, St. Andrews, KY16 9SS, UK}

I \author{V.Archontis}
\affil{School of Mathematics and Statistics, St. Andrews University, St. Andrews, KY16 9SS, UK}
\affil{Department of Physics, University of Ioannina, 45110, Ioannina, Greece}

 \begin{abstract}
 We have performed 3D MHD simulations to study the effect of partial ionization in the process
of magnetic flux emergence in the Sun. In fact, we continue previous work and we now focus: 1) on the emergence
of the magnetic fields above the solar photosphere and 2) on the eruptive activity, which follows the
emergence into the corona. We find that in the simulations with partial ionization (PI), the structure
of the emerging field consists of arch-like fieldlines with very little twist since the axis of the
initial rising field remains below the photosphere. The plasma inside the emerging volume is less dense and
it is moving faster compared to the fully ionized (FI) simulation. In both cases, new flux ropes (FR) are
formed due to reconnection between emerging fieldlines, and they eventually erupt in an ejective manner towards the
outer solar atmosphere. We are witnessing three major eruptions in both simulations. At least for the first eruption,
the formation of the eruptive FR occurs in the low atmosphere in the FI case and at coronal heights in the PI case.
Also, in the first PI eruption, part of the eruptive FR carries neutrals in the high atmosphere, for a short
period of time.
Overall, the eruptions are relatively faster in the PI case, while a considerable amount of axial flux is found above the
photosphere during the eruptions in both simulations.

 \end{abstract}


\section{Introduction} \label{sec:intro}

Studying the process of magnetic flux emergence in the Sun is 
important towards a better understanding of the solar magnetic activity. 
Observations and simulations have shown that the emergence of magnetic 
flux can lead to the formation of active regions, which can spawn 
different types of solar activity, such as flares, jets and eruptions that 
can release vast amounts of energy and charged particles into the solar system, influencing space weather conditions \citep[e.g.,][]{Demoulin_etal2002,Nindos_2003ApJ...594.1033N,Innes_etal2010,Raouafi_etal2010,Hong_etal2011,Patsourakos_etal2013}. One important property of the plasma in the low solar atmosphere is that 
it has a low degree of ionization, and thus, the presence of neutrals can 
affect the emergence of the magnetic field and the solar dynamics at the solar atmosphere and above. The Solar Optical Telescope (SOT) onboard Hinode \citep{hinode_2007SoPh..243....3K,SOT_2008SoPh..249..167T} has provided detailed observations of solar dynamics, revealing phenomena influenced by partial ionization. Chromospheric jets and Ellerman bombs, linked to magnetic reconnection, highlight the role of neutral-ion interactions in the upper photosphere and lower chromosphere \citep[e.g.,][]{Matsumoto2008PASJ...60..577M,Singh_2012ApJ...759...33S}. Partial ionization also impacts the propagation and damping of transverse waves, potential key drivers of coronal heating, through mechanisms like perpendicular resistivity \citep[e.g.,][]{Soler2009ApJ...707..662S,Singh2010NewA...15..119S}. Additionally, the SOT has observed magnetic Rayleigh-Taylor instabilities in prominence bubbles and plumes, processes shaped by partial ionization through altered instability growth rates \citep[e.g.,][]{Arber_etal2007,Berger2011Natur.472..197B}. These findings underscore the importance of partial ionization in solar atmospheric dynamics. 

Moreover, over the past years, 3D numerical simulations using fully ionized 
plasma \citep[e.g.,][]{Fan_2001,Manchester_etal2004,Archontis_Torok2008,Archontis_etal2012,Leake2022ApJ...934...10L} showed that the emergence of a twisted flux tube  rising from beneath the solar surface can lead to the formation of a FR through shearing and reconnection of field lines at the lower atmosphere. The rising motion of the FR may evolve into an ejective or a confined eruption. These eruptions may account for the ejections of filaments and CMEs \citep{Archontis_etal2012,Moreno-Insertis_etal2013,Syntelis_etal2017}. 
\par In the present study we are using very similar numerical simulations, for our fully ionized case, with the simulations by \citet{Syntelis_etal2017} who showed that the emergence of the twisted flux tube into an unmagnetized corona can lead to recurrent eruptions. Those simulations showed that a new FR is formed above the polarity inversion line (PIL) due to reconnection of low-lying sheared fieldlines. Once formed, the FR undergoes an instability, which together with tether-cutting reconnection in the current sheet below the FR leads to an ejective eruption. In this paper, we focus more on the characteristics and properties of the first eruption while we study the time evolution of energies and flux for all the subsequent eruptions and we compare the partially ionized case with the fully ionized case.

\par In most of the afore-mentioned studies the plasma is considered fully ionized. The inclusion of partial ionization in the MHD framework is achieved through the generalized Ohm's law \citep{1965RvPP....1..205B}, which has been implemented in single-fluid MHD simulations by various studies \citep[e.g.,][]{Brandenburg1994ApJ...427L..91B, Osullivan2007MNRAS.376.1648O, Cheung2012ApJ...750....6C,Martinez-Sykora_etal2012, Gonzalez2018A&A...615A..67G,Bifrost_2020A&A...638A..79N,Muram_2021ApJ...923...79R}. Additionally, there are studies that investigate the small-scale effects of ion-neutral interactions using multi-fluid MHD approaches \citep[e.g.,][]{Zaqarashvili2011A&A...529A..82Z, Leake2012ApJ...760..109L, Khomenko2014PhPl...21i2901K}.
. However, there are a few numerical studies with magnetic flux emergence including a partially ionized plasma \citep{Leake_etal2005,Arber_etal2007,Leake_etal2013b,Nobrega2020A&A...633A..66N}.  \citet{Leake_etal2013b} performed 2.5D numerical simulations including partially ionized plasma and they reported that the development of unstable coronal structures is unlikely to occur. In a recent work, \citet{Chouliaras_2023ApJ...952...21C}, hereinafter referred to as Paper I, showed that in 3D numerical simulations of magnetic flux emergence with partially ionized plasma, ejective solar eruptions are possible.

\par In Paper I, our study focused on the effect of partially ionized plasma on the emerging field below and at the solar surface. We showed that the slippage of ions through the plasma close to the photosphere changes the emerged magnetic field structure, which in turn affects the properties of the emerged plasma above the photosphere. For instance, the FI apex brings more dense plasma in the solar surface in comparison to the PI simulations.

\par In the present paper we mainly study the emergence above the solar surface. More specifically, we investigate the impact of PI on the dynamics and physical properties of both the emerged and erupting plasma. The paper is organized as folows: Section 2 is a short description of our numerical setup. Section 3 presents the main results of our simulations. In section 4 we present the conclusions of our study.


\section{Numerical setup}
We performed 3D numerical simulations and numerically solved the 3D time-dependent compressible resistive MHD equations. For the numerical integration of the complete system of MHD equations, we employed the Lare3D code (see \citealp{Arber_etal2001}). The numerical setup is thoroughly explained in Paper I, but we briefly discuss some key aspects here.

The computational domain has a physical size of $64.8^3 \mathrm{Mm}$ in a $420^3$ uniform grid. The interior extends from $-7.2$~Mm$\le z < 0$~Mm, the photospheric-chromospheric layer from $0$~Mm$\le z < 1.8$~Mm, the transition region from $1.8$~Mm$\le z < 3.2$~Mm, and an isothermal corona from $3.2$~Mm$\le z < 57.6$~Mm. We assume periodic boundary conditions in the $y$ direction. Open boundary conditions are used in the $x$ direction and at the top of the numerical box, while the bottom boundary is set to be closed. 

We utilized two models in this study. One assumes a fully ionized plasma, while the other considers a partially ionized plasma. We modified the MHD equations to incorporate the effect of partial ionization in both the magnetic field evolution and the equation of state. Both flux tubes begin with the same plasma beta, but due to the different equations solved for the stratification of the atmosphere (details in Paper I), they have initial magnetic fields with distinct magnitudes: $B_{0{PI}}=7882G$ and $B_{0{FI}}=3150G$, respectively. 
\begin{figure*} 
\centering
\includegraphics[width=1.1\textwidth]{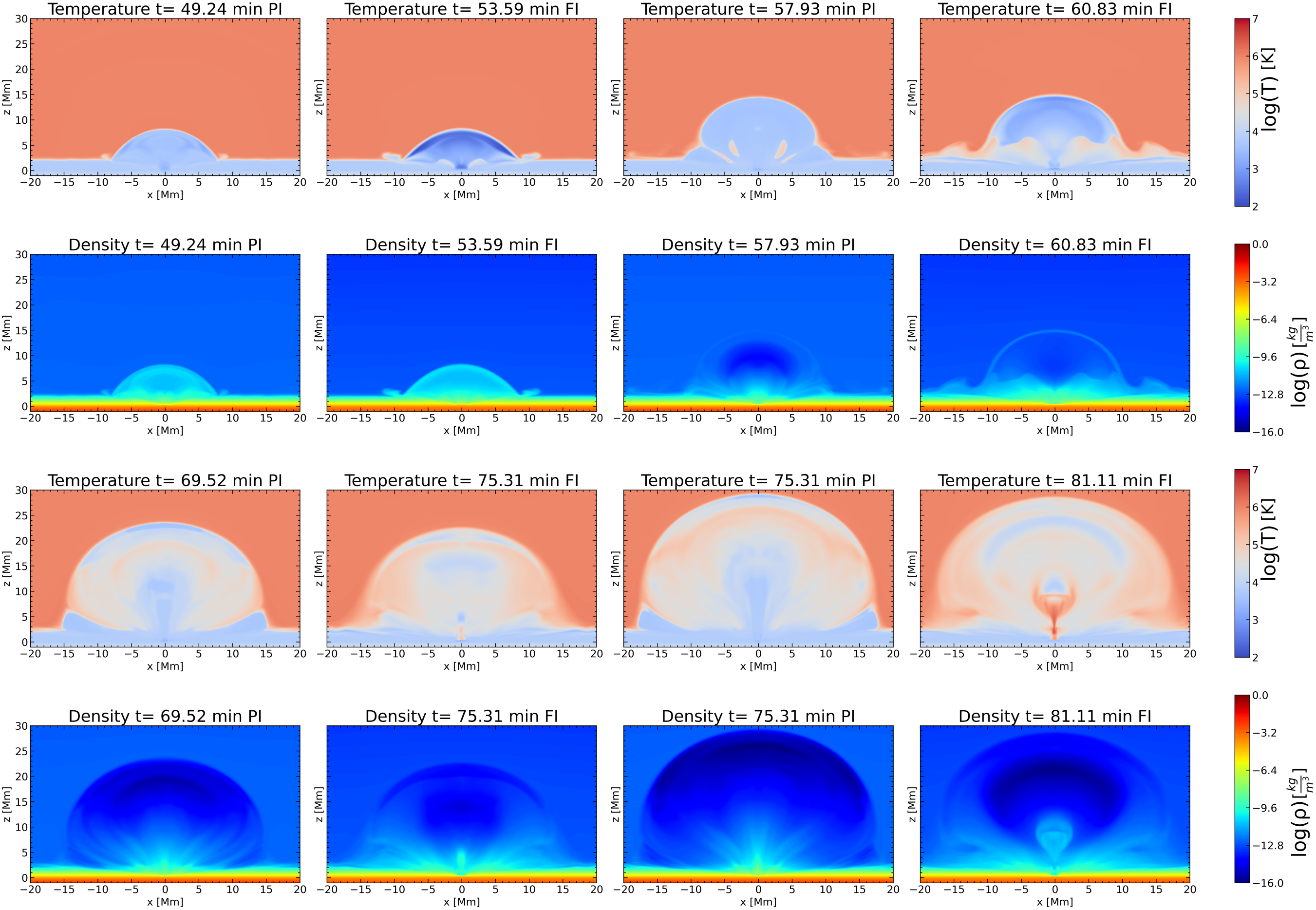}
\caption{Contour plots depicting the logarithm of temperature and density at four different times for both simulations. The plots represent the early evolution of the magnetic loop in the solar atmosphere at the xz midplane. At each of the four times, the apex of the emerging field is located at the same height in both simulations.}
\label{fig:pre1}
\end{figure*}

\begin{figure*}[htbp]
    \centering
    \includegraphics[width=0.49\textwidth]{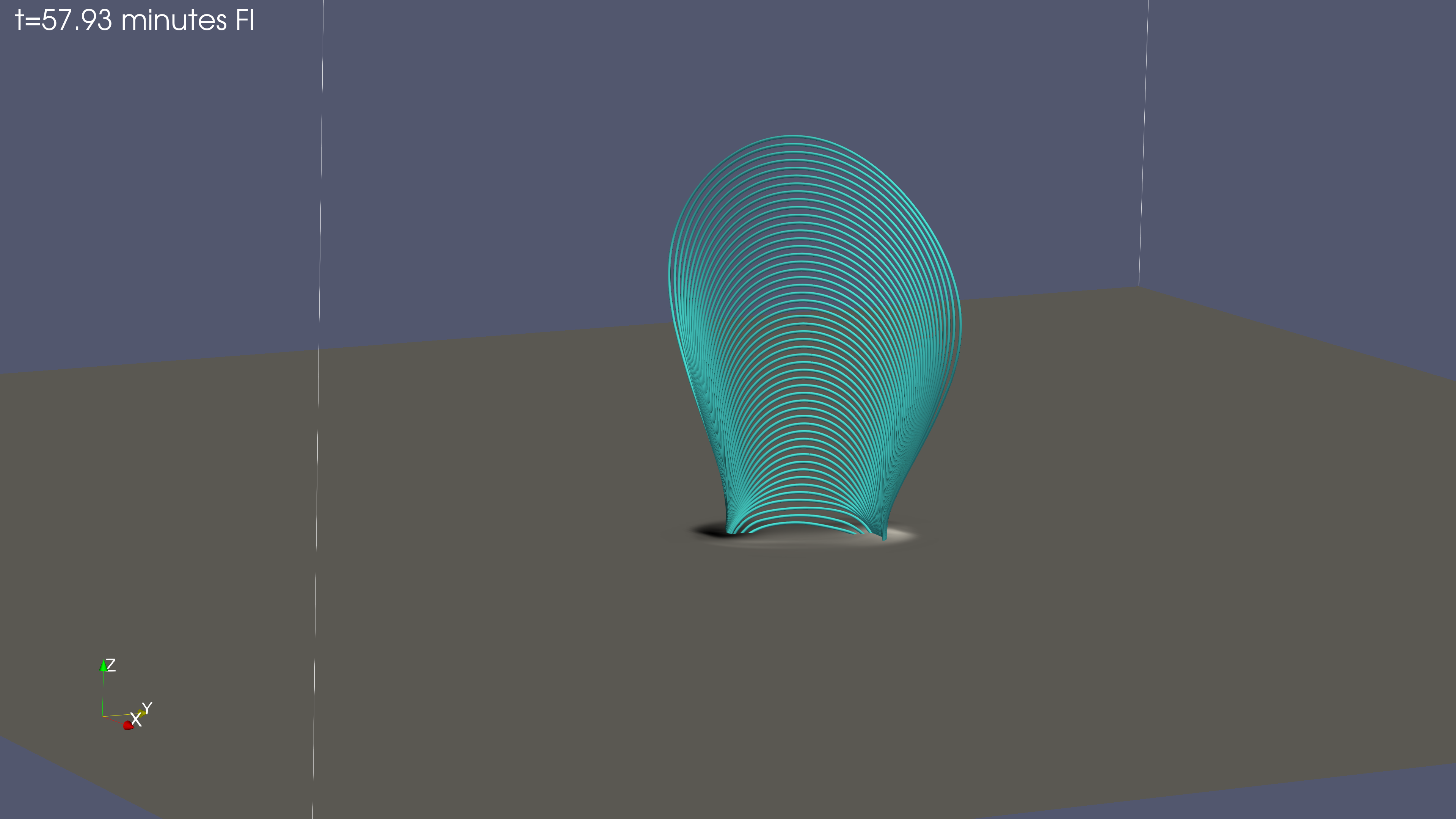}
    \hfill
    \includegraphics[width=0.49\textwidth]{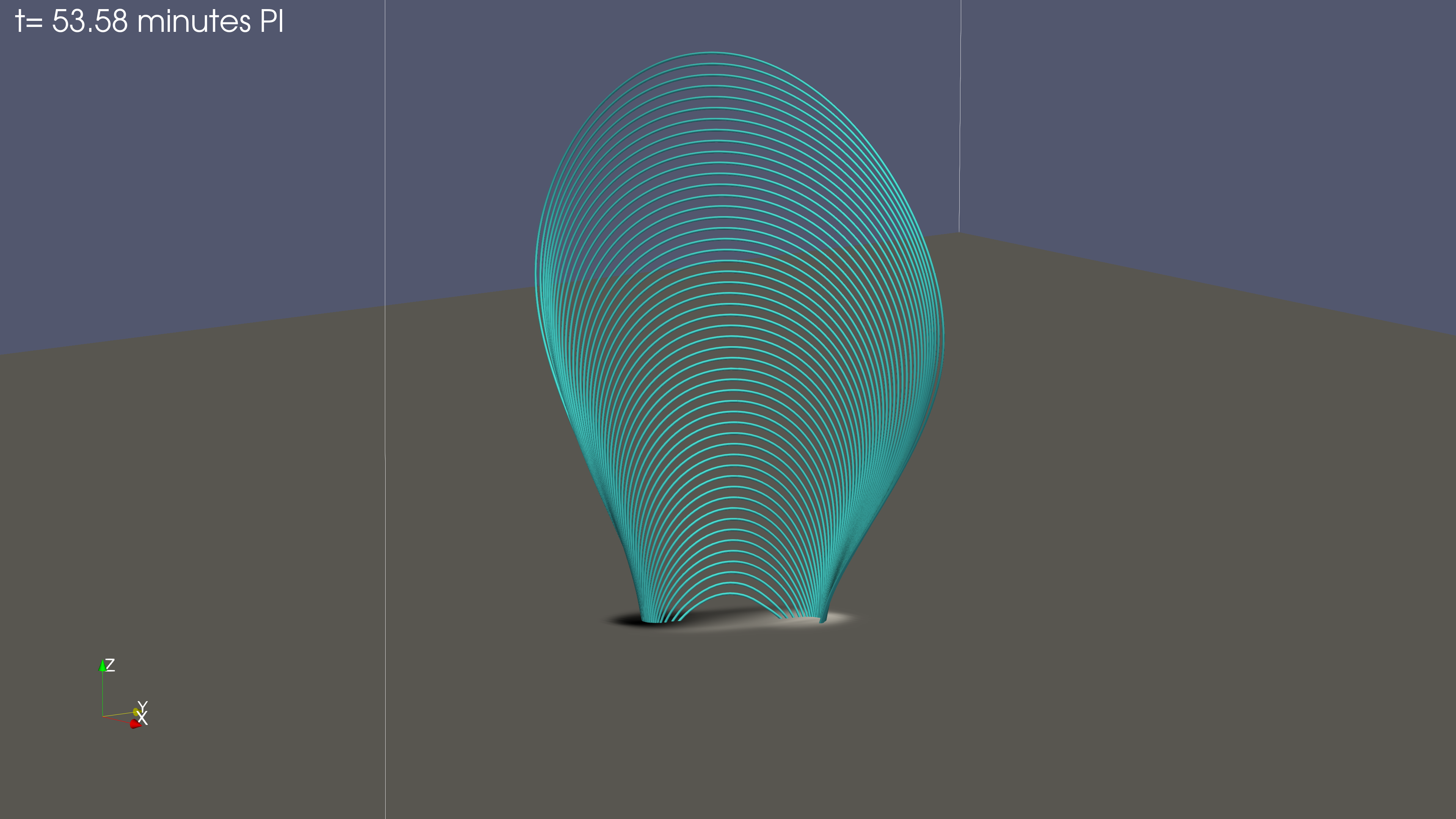}
    
    \includegraphics[width=0.49\textwidth]{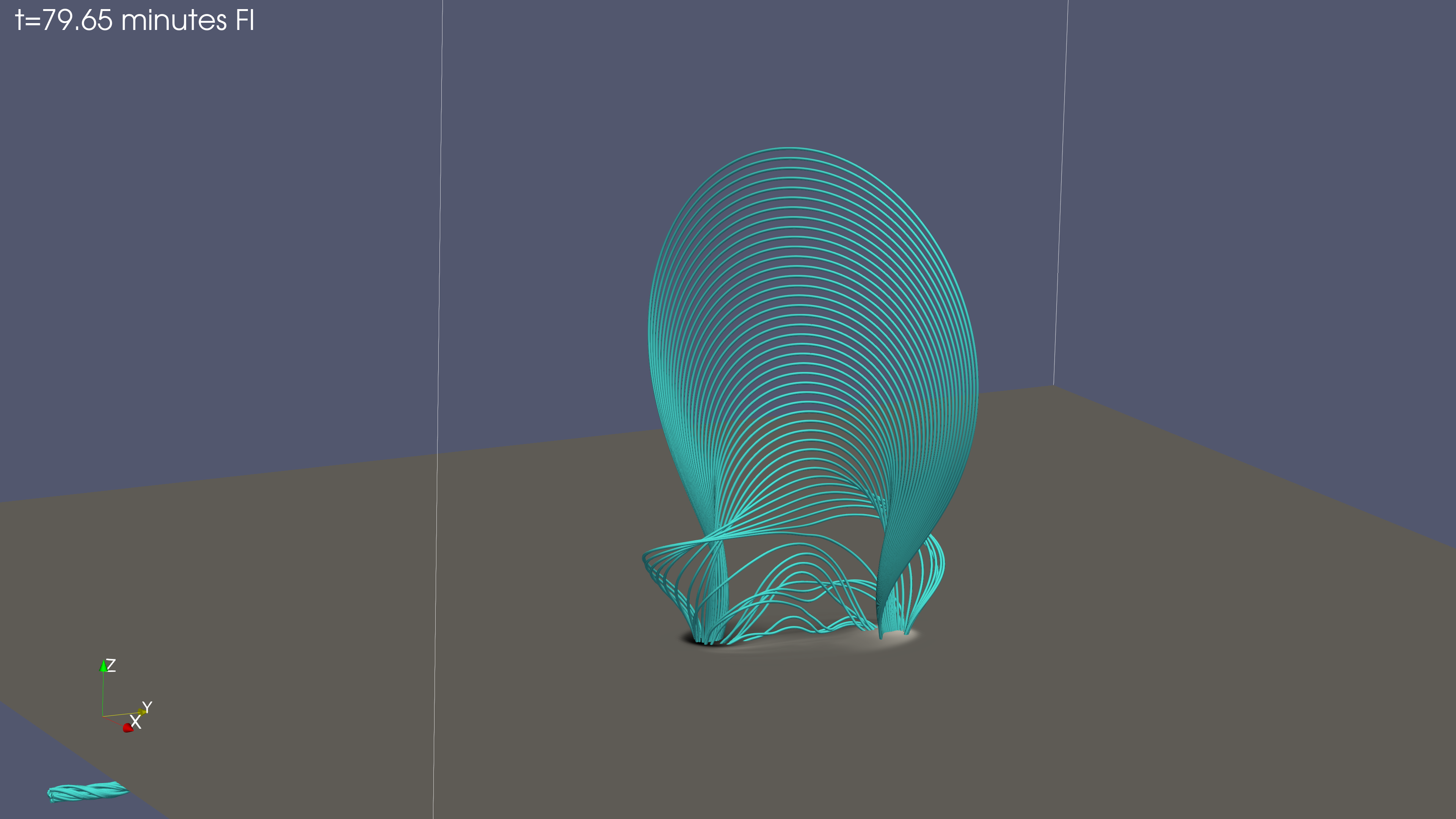}
    \hfill
    \includegraphics[width=0.49\textwidth]{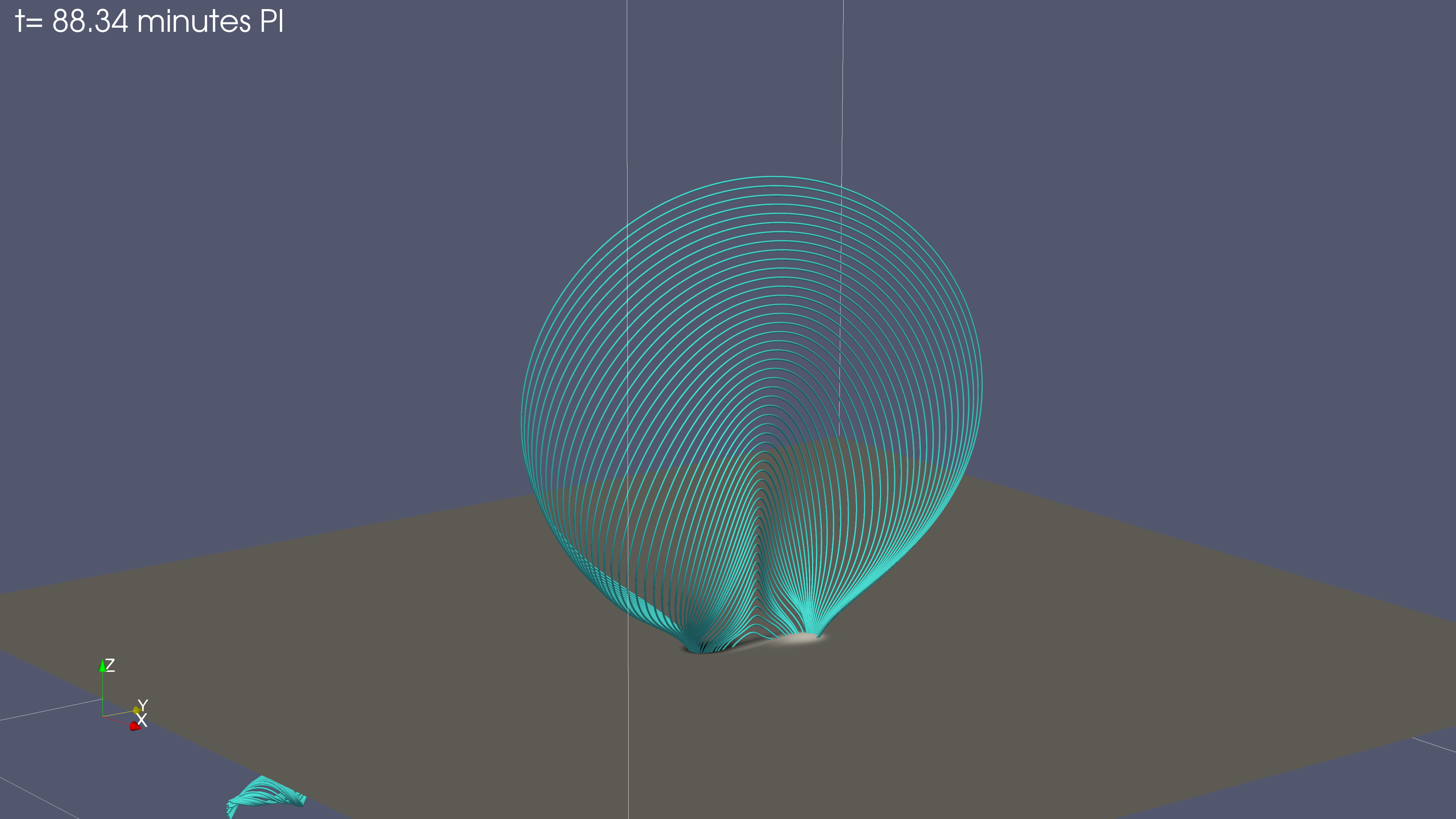}
    
    \includegraphics[width=0.49\textwidth]{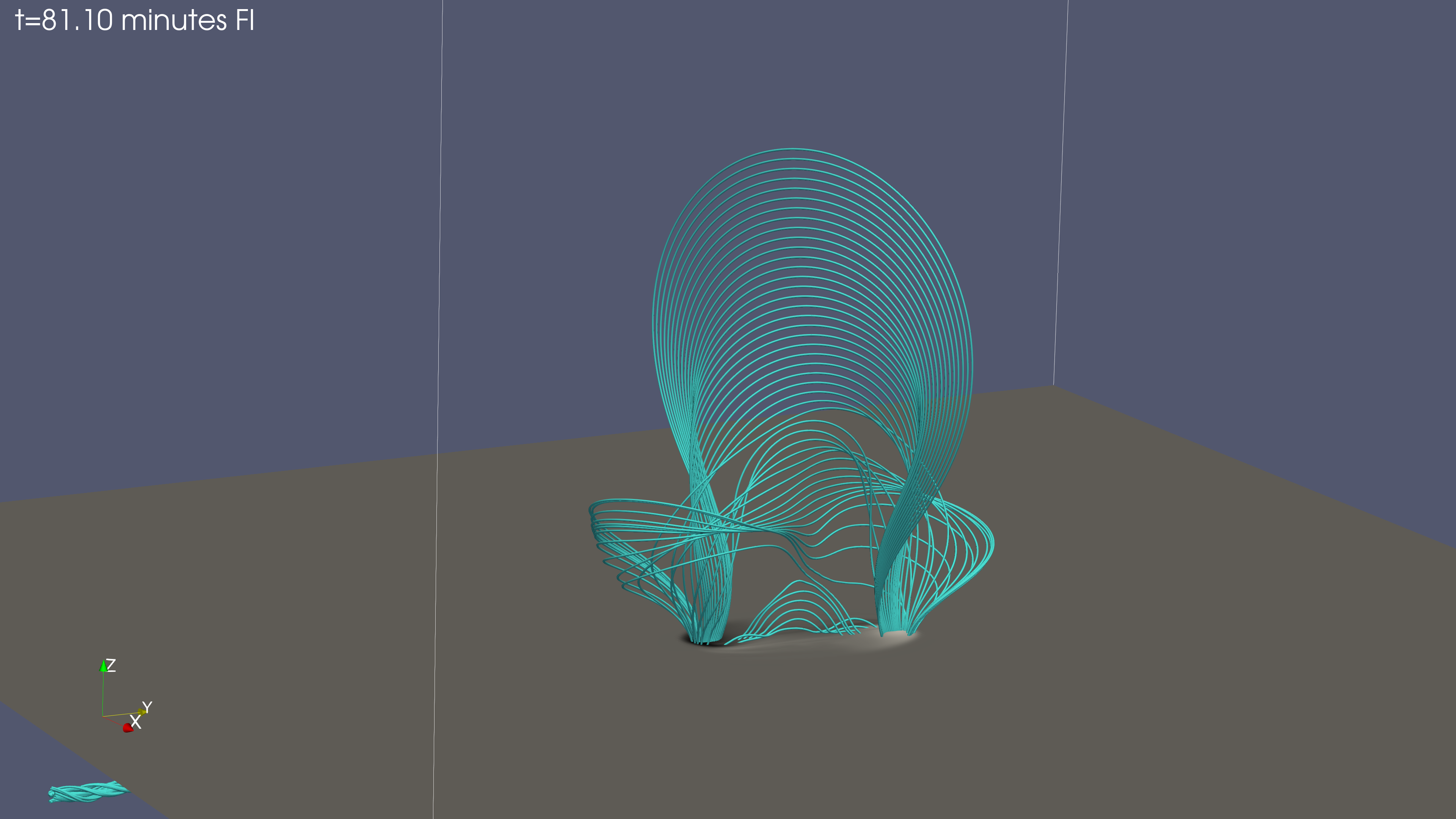}
    \hfill
    \includegraphics[width=0.49\textwidth]{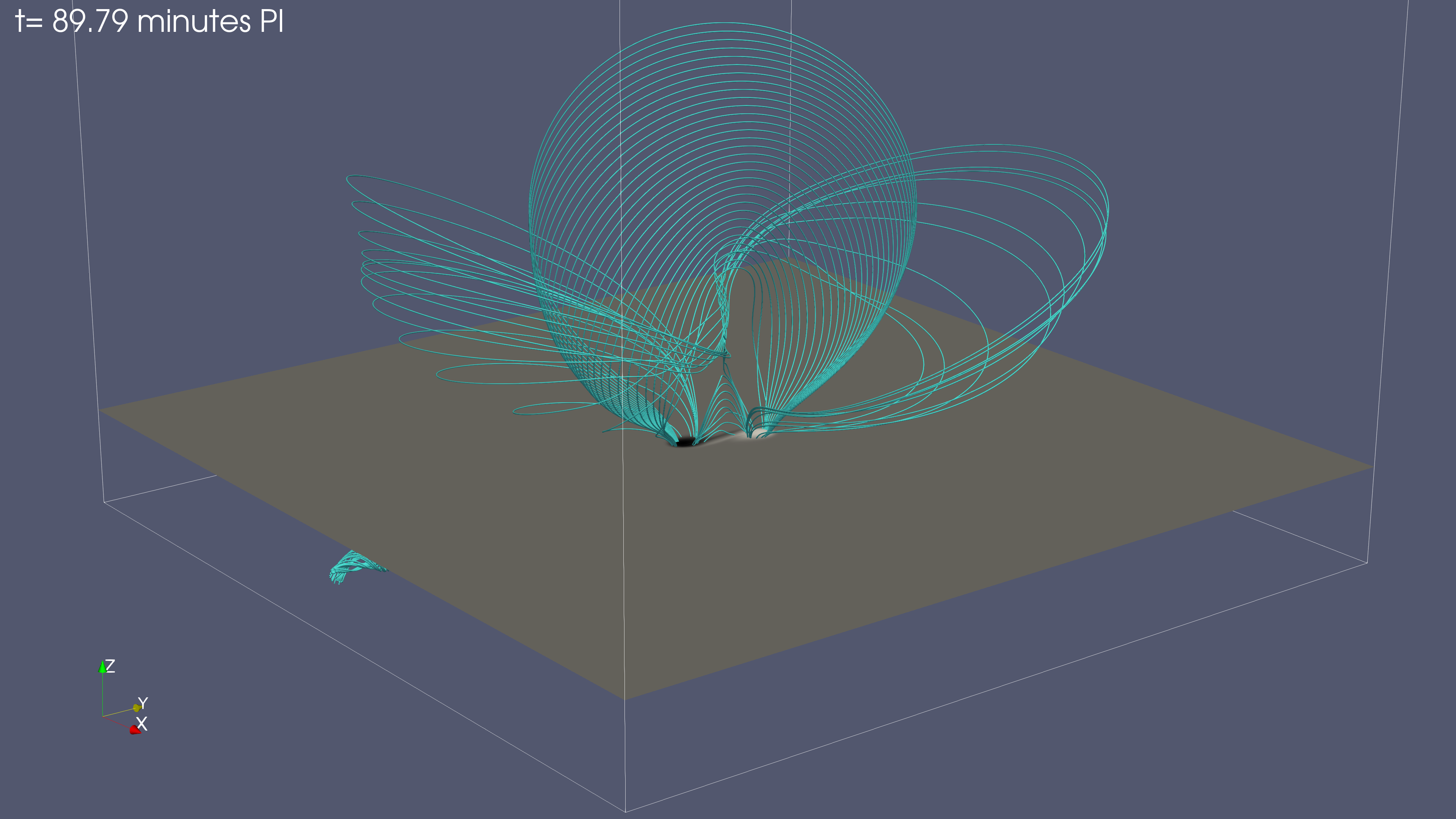}
    
    \caption{3D magnetic field line topology for both simulations with traced magnetic field lines along a vertical line at the center of the box.}
    \label{fig:blue_trace}
\end{figure*}

\begin{figure*}[htbp]
    \centering
    \begin{minipage}{.490\textwidth}
        \centering
        \includegraphics[width=\textwidth]{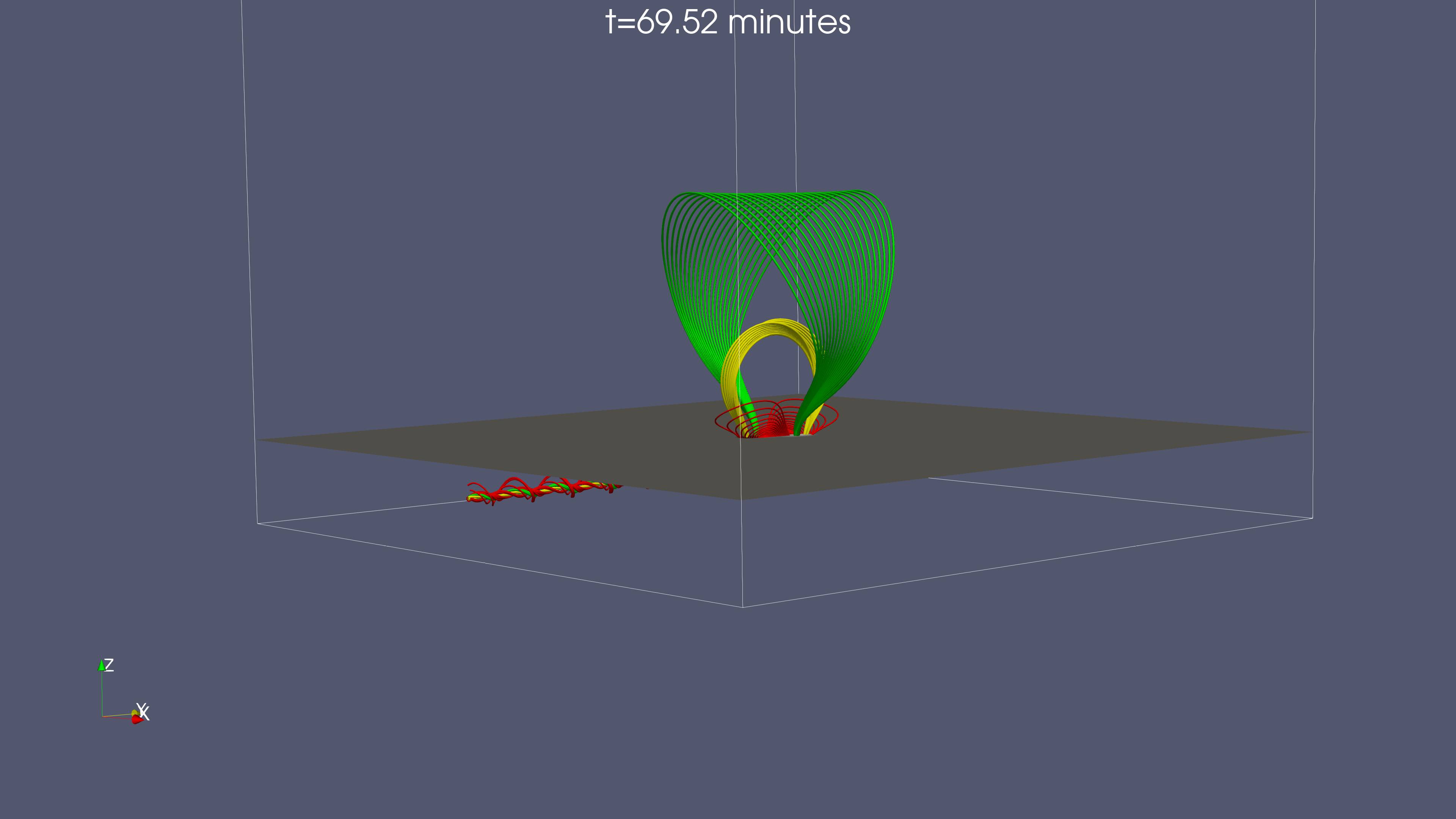}
        \parbox{0pt}{\hspace*{\fill}(a)\hspace*{\fill}}
    \end{minipage}
    \hfill
    \begin{minipage}{.490\textwidth}
        \centering
        \includegraphics[width=\textwidth]{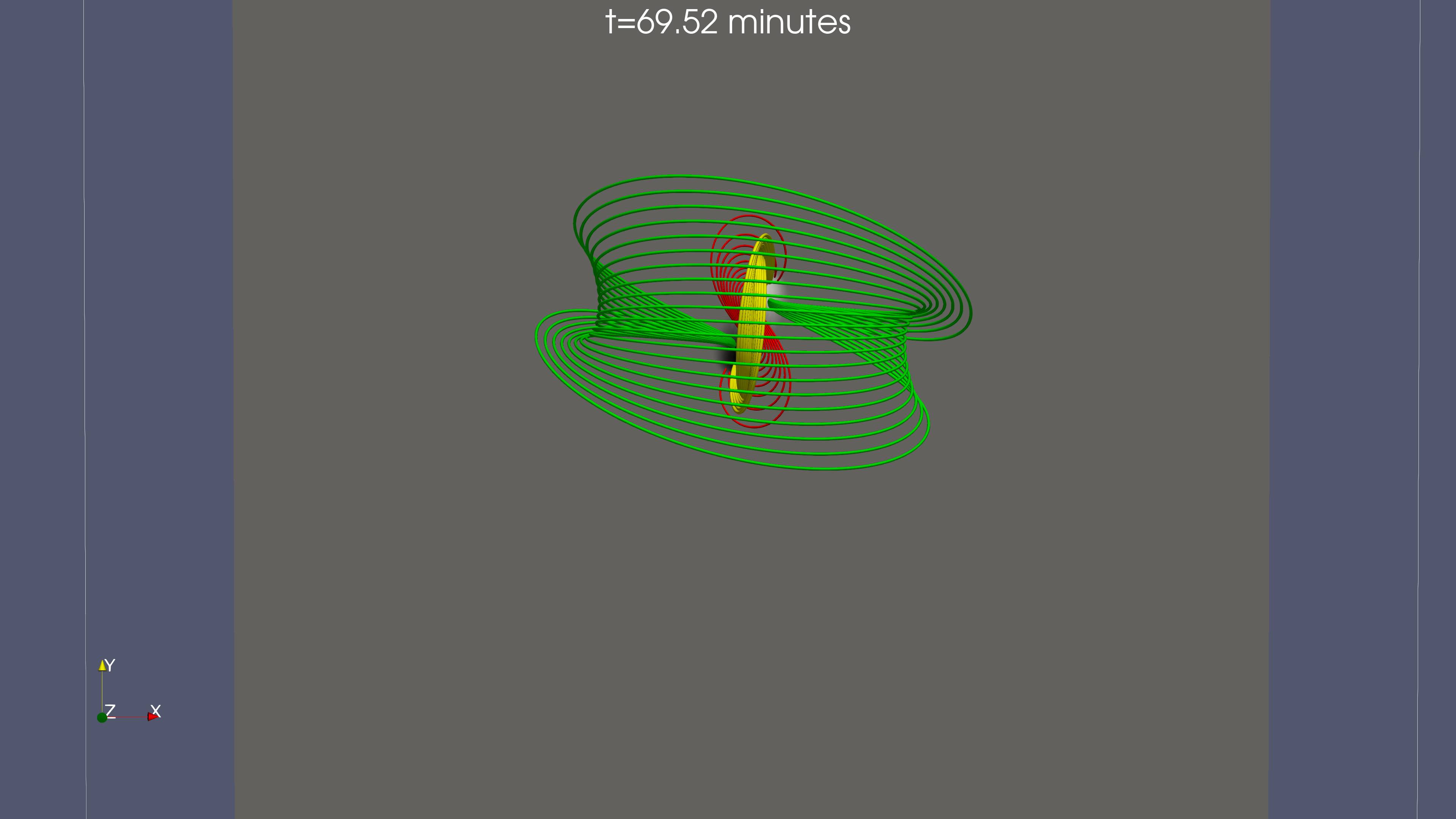}
        \parbox{0pt}{\hspace*{\fill}(b)\hspace*{\fill}}
    \end{minipage}
    
    \begin{minipage}{.490\textwidth}
        \centering
        \includegraphics[width=\textwidth]{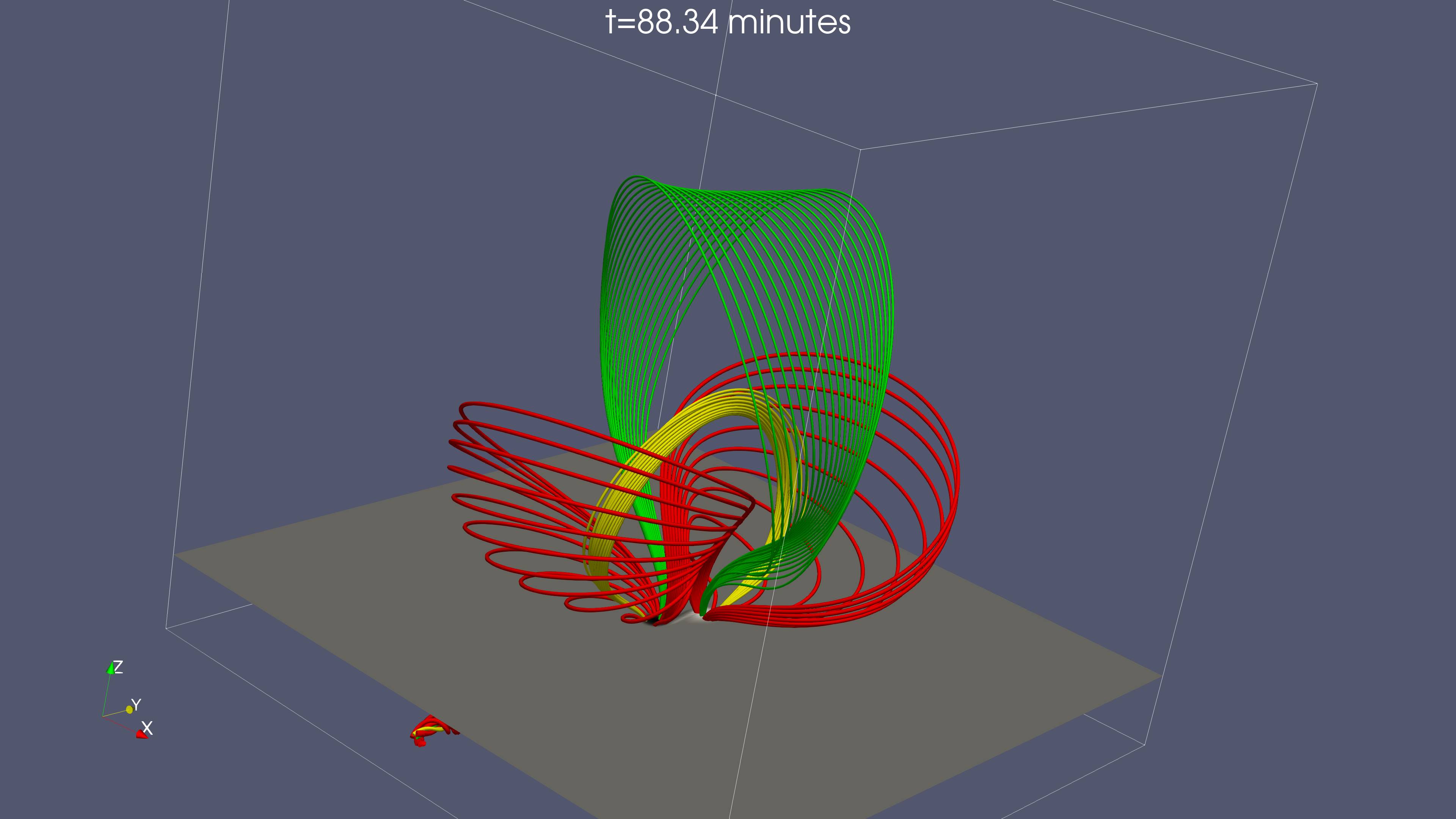}
        \parbox{0pt}{\hspace*{\fill}(c)\hspace*{\fill}}
    \end{minipage}
    \hfill
    \begin{minipage}{.490\textwidth}
        \centering
        \includegraphics[width=\textwidth]{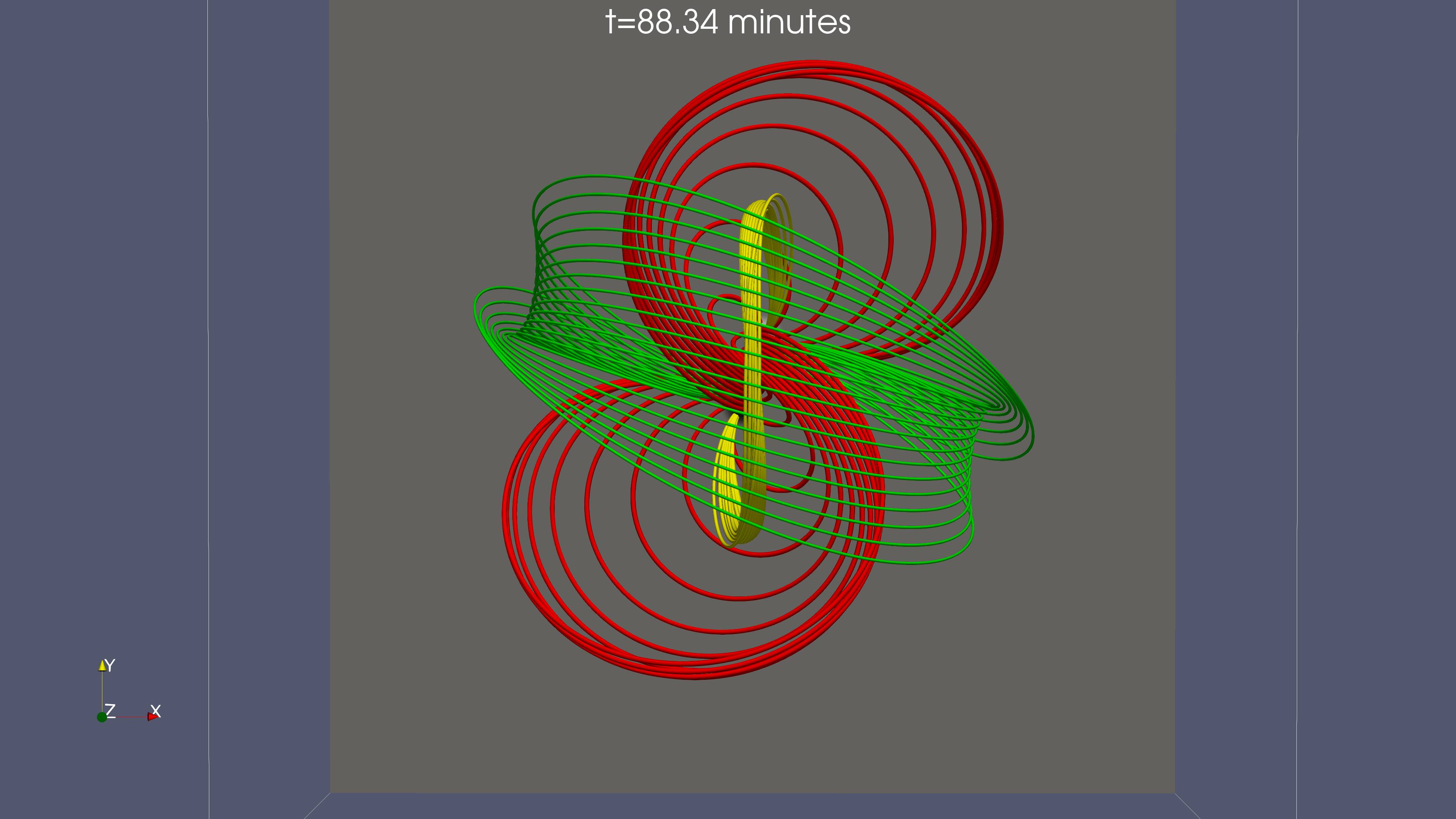}
        \parbox{0pt}{\hspace*{\fill}(d)\hspace*{\fill}}
    \end{minipage}

    \begin{minipage}{.490\textwidth}
        \centering
        \includegraphics[width=\textwidth]{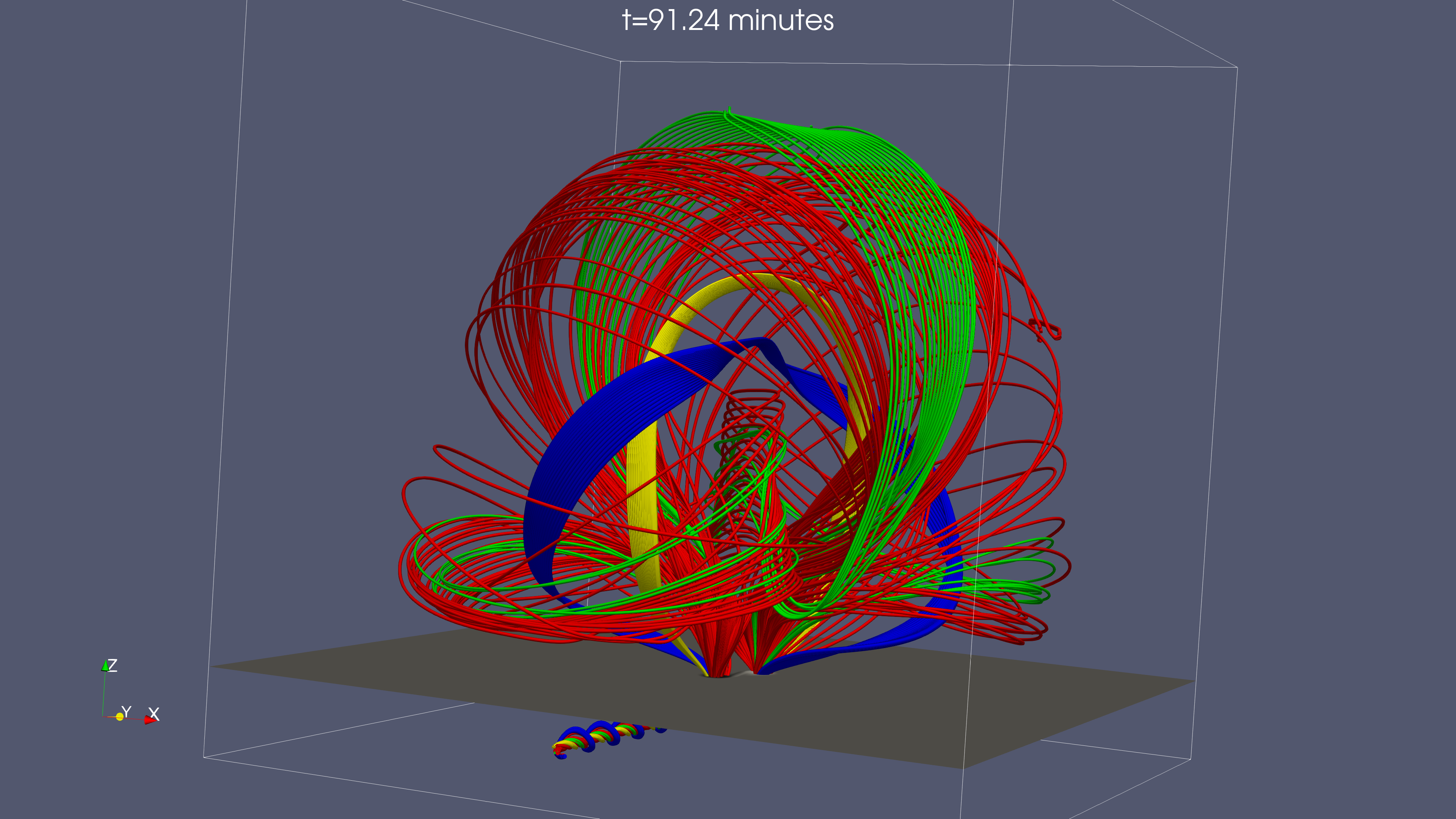}
        \parbox{0pt}{\hspace*{\fill}(e)\hspace*{\fill}}
    \end{minipage}
    \hfill
    \begin{minipage}{.490\textwidth}
        \centering
        \includegraphics[width=\textwidth]{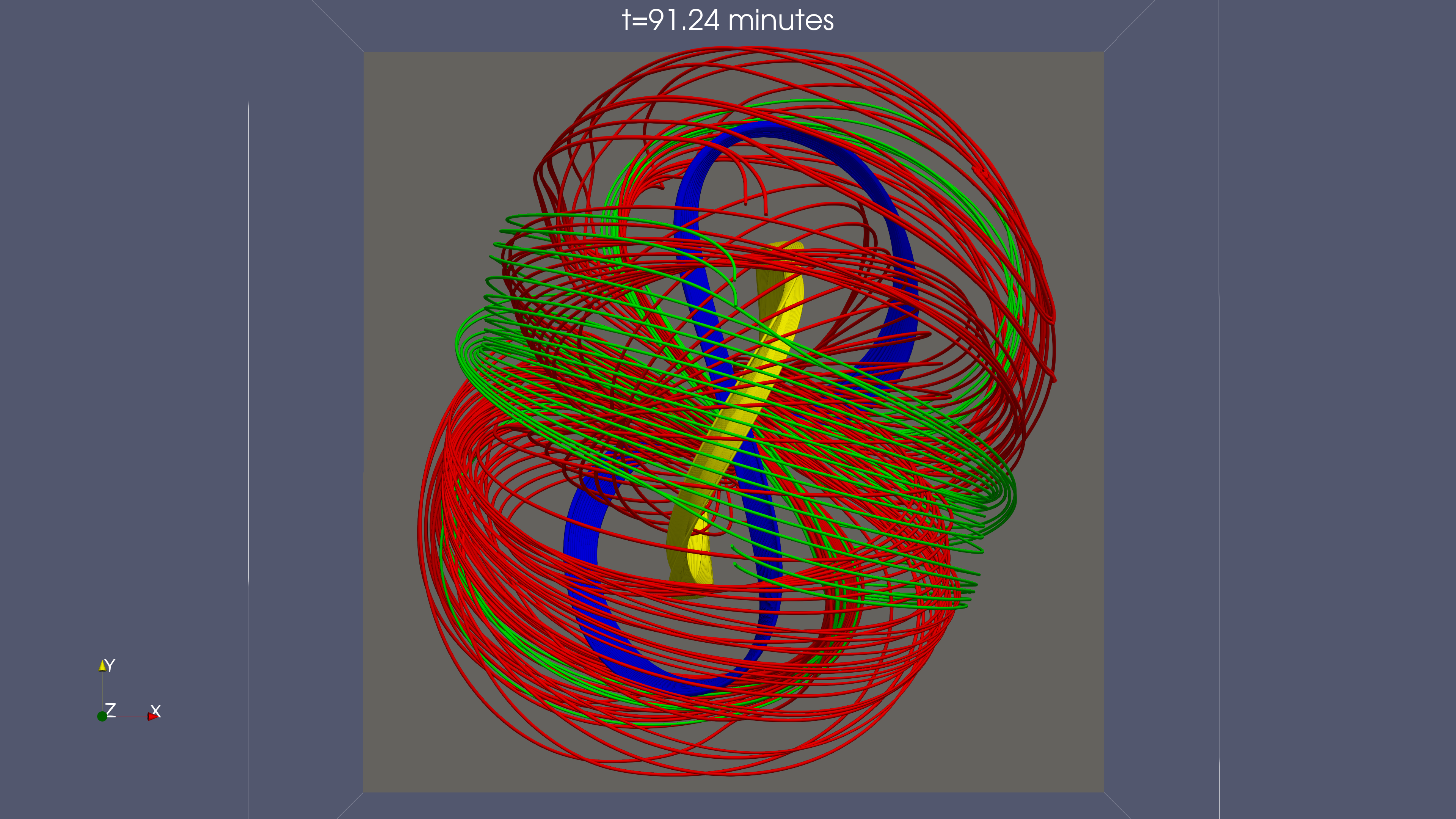}
        \parbox{0pt}{\hspace*{\fill}(f)\hspace*{\fill}}
    \end{minipage}
    
    \caption{Field line morphology of the first eruption in PI simulation at (a,b) t=69.52 minutes side and top view (c,d) t=88.34 minutes side and top view and (e,f) t=91.24 minutes side and top view.}
    \label{fig:morph}
\end{figure*}

\vspace{1cm} 

\section{Results}
 In Paper I, we have demonstrated that both flux tubes reach the solar surface at approximately $t=25$ minutes. The magnetic flux remains there until the Acheson criterion (\citet{Acheson1979}) is satisfied.  In the PI simulation, the first magnetic flux elements emerge toward the solar atmosphere 15 minutes after their emergence into the photosphere. Conversely, the magnetic flux in the FI simulation starts to emerge 20 minutes after its emergence to the solar surface, due to the denser plasma it carries, which slows its rise. In both cases the emergence at the photosphere is followed by the formation of bipolar region with PIL in between 
the opposite polarities. 
\subsection{The pre-eruptive phase}
Our study in Paper I showed that after the expansion of the emerging field into the solar atmosphere, a number of eruptions occured above the PILs of the bipolar regions. In this section, we study the pre-eruptive phase in the two numerical experiments. In Figure \ref{fig:pre1}, we show the evolution of various plasma properties for the two simulations, FI and PI. The first two rows correspond to the evolution of the emerging field when its apex has reached a height of around 8 Mm and up to the evolution of the plasma when it has expanded well into the corona, reaching a height of around 15 Mm. The temperature distribution (first row) shows that the interior of the PI loop is less cool compared to the FI loop. This is due to the presence of neutrals at the interior of the loop. The interaction with the ions through collisions prevents the plasma from intense adiabatic cooling, which is apparent in the FI case (e.g. $t=53.59$ minutes, top,second column). In the second row, we show the density distribution, at these two stages of the evolution. 
As we have shown in Paper I, the FI emergence brings more dense plasma above the photosphere. This is visible at the first stage ($t=49.24 $ minutes, $t=53.59$ minutes) and it becomes more apparent as the field expands higher in the atmosphere ($t=57.93$ minutes, $t=60.83$ minutes). In the third and fourth rows of Figure \ref{fig:pre1}, we show the temperature and density distributions at later stages of the evolution. In both cases, we find that cool and dense plasma, which originates at the center and inside the emerging field, is moving upwards. In the FI case, this is related to the formation of a new FR at the lower atmosphere, as a result of reconnection between sheared fieldlines, which can bring dense plasma in the higher atmosphere and lead to eruptions of filament-like structures, as it has been shown in previous studies \citep{vanBallegooijen_etal1989,Magara_etal2001,Archontis_Torok2008,Fan_2009,Leake_etal2014,Syntelis_etal2017}. In the PI case, there is no formation of a new FR at this stage of the evolution. The dense material is carried upwards by emerging arch-like fieldlines. We show the structure of the emerging field in more detail in the next subsection.
Now, at the times where the apex of the emerging field has reached the height of about 22Mm, the dense volume inside the emerging field in the PI case has expanded more, vertically and horizontally. In the FI case, the dense and cool material remains at $2-4$ Mm above the photosphere. The temperature distribution shows that in the PI case, there is no heating underneath the expanding flux system ($t=69.52$ minutes), while in the FI case there is heating at the central region ($x=0$, $y=0$), exactly under the new FR. This heating becomes much more apparent at $t=81.11$ minutes, where it is concentrated within a Y-shaped area under the cool and dense rising plasma. Still, at this time, there is no profound heating  in the PI case. This could indicate that there are two important differences between the
FI and the PI case. Firstly, the structure of the magnetic field inside the emerging volume is different between the two cases and secondly, reconnection at the low atmosphere occurs only in the FI case, and it heats the plasma localy. In the following section, we will ellaborate on the afore-mentioned differences. Now, looking again at the temperature distribution, but at the area between the inside rising field and the apex of the emerging field, we find that the local plasma temperature is higher in the PI case. Also, at this stage of the evolution, the plasma density is higher inside the rising field in the FI case (e.g. last panel, fourth row) in comparison to the PI case (e.g. 3rd panel, fourth row). In principle, this is consistent with the results of  Paper I based on the slippage effect. In section 3.4, we will show that, at a later stage of the evolution, the inside rising field erupts in an ejective manner in both cases.
\subsection{Topology of fieldlines and height-time profile.}
To understand better the inside structure of the magnetic field, which emerges above the photosphere we trace fieldlines at three different times, until just before the first eruption. This is shown in Figure \ref{fig:blue_trace}. In all snapshots, we trace fieldlines from the central vertical line at (x,y)=(0,0), from the photosphere until z=30Mm. The FI case is shown in the first column and the PI case in the second column in Figure \ref{fig:blue_trace}. Here, it is important to mention that our previous study (Paper I) showed that the axis of the twisted emerging flux tube in FI rises above the photosphere and it stays within the photosphere, during the whole evolution of the system. However, in the PI case, the axis of the emerging field remains below the photosphere during the simulation. This result is consistent with the results by \citet{Leake_etal2013b} in a similar numerical experiment.

In the first row, the fieldlines show the emergence of the field into the corona,
before the formation of a new FR. The horizontal slice shows the distribution of Bz at the photosphere (white shows positive and black shows negative polarity). A clear difference between the two cases is the shape of the fieldlines, which all together form the 3D structure of the field. At the low atmosphere, in the FI case the fieldlines have more horizontal orientation, connecting the two polarities. These fieldlines are part of the field close to the axis of the emerging field, which has risen above the photosphere. At larger heights the fieldlines form the expanding volume with the fan-like shape, which has been shown in many previous studies. In the PI case, the low-lying fieldlines above the PIL, have less horizontal orientation and they are more stretched upwards, forming an arcade-like structure at all heights, connecting the positive with the negative polarity of the emerging bipole. This is because the axis of the emerging twisted flux tube is below the photosphere and, thus, what we see in this panel is the fieldlines above the axis of the flux tube that they have managed to emerge above the photosphere.

In the second row, we show the fieldlines  before the first  eruption. In the FI case, there is a number of twisted fieldlines above the PIL, which are formed by reconnection of sheared fieldlines as it has been described in detail in previous studies \citep{Manchester_etal2004,Archontis_Torok2008,Syntelis_etal2017,Leake2022ApJ...934...10L}. These new fieldlines form the FR, which is oriented along the y-direction and it will erupt at a later time. In the PI case, the field continues to expand and the middle part shows an upward stretching as their footpoints connect down to the photosphere without experiencing any reconnection. 
The upward stretching of the field lines is due to the fact that the plasma behind the apex of the emerging field is moving faster than the apex itself, causing vertical stretching of the field lines at the center, we will study this in more detail in subsection 3.3. At this stage of the evolution it is clear that the structure of the emerging field is substancially different in the two cases. In the FI case a FR has already been formed and it rises upwards while in the PI case the magnetic field adopts an arcade like configuration.
\par In the third row, the FR in the FI case has reached coronal heights and it has started to erupt with higher speed. In the PI case we are witnessing for the first time reconnection between field lines which have expanded at the two side flanks of the emerging field. These field lines reconnect at the center of the emerging volume (around $x=0,y=0$) at coronal heights. Reconnection between similar field lines continue to occur at progressively larger heights leading to the formation of a FR which eventually will erupt in an ejective manner. 
The mechanism for the formation of the FR in the FI case is similar to the formation which has been described in previous similar experiments  \citep[e.g.,][]{Manchester_etal2004,Archontis_Torok2008,Syntelis_etal2017,Leake2022ApJ...934...10L}. To study the formation of the FR in the PI case we show the evolution of the magnetic field lines from various locations in the 3D space during the emergence of the field.
\par Figure \ref{fig:morph} is a visualization of 3D magnetic field lines at three different times (top row t=69.52 minutes middle row t=88.34 minutes and bottom row t=91.24 minutes). At the top row, green field lines represent the apex of the emerging field. The yellow field lines have been traced from the vertical xz mid-plane from a height where the magnetic field is oriented preferably along the y direction and above and below, the Bx component is changing sign, from negative to positive. Red field lines are traced from the photosphere, around the center x=0,y=0 from the two opposite polarity sides of the PIL. The red field lines are not connected to each other at this stage of the evolution. The top view (top right panel) shows that each set of the red field lines have a J-like shape. The middle row show field lines traced for similar locations at a later times. We have to highlight that these are not the same field lines as the ones visualized at the earlier stage (top row). However they show the overall expansion of the field and the J-like field lines which have not been reconnected. As time goes on, the vertical segments of the Js come closer
together and a strong current sheet is build up, where reconnection
will start soon. At the bottom row of Figure \ref{fig:morph} we show the field line topology after reconnection between the J-like field lines has occured. The blue field lines are the new field lines which have been formed after the J-like field lines have reconnected and they have an overall S-like shape (top view). Now, the yellow field lines have been traced from a height between the new reconnected field lines (blue) and the apex of the emerging field (green). At this stage of the evolution a new FR is formed, which will also erupt (as in the FI case) into the outer atmosphere. The core of the new FR consists of the blue field lines. Above and behind the core of the FR there are field lines twisted around it. For example some of the low-lying red field lines are wrapped around the flanks of the core field lines of the FR. Notice that the reconnected field lines (blue) in the bottom row in this Figure \ref{fig:morph} are basically the same with the reconnected field lines which are shown at the bottom right panel in Figure \ref{fig:blue_trace}. 
We have not studied the exact mechanism for the origin of the eruption(s) in the PI case but we anticipate that reconnection between the Js and the tethers of the envelope fieldlines, such as the green fiedlines in Figure \ref{fig:morph}, releases the downward tension of the enevelope field and it contributes to the onset of the eruption. The mechanism for the eruptions in the FI case is very similar to that described in previous studies e.g. \citet{Syntelis_etal2017}.
\par
\begin{figure}[!ht]
\centering
\hspace*{-1cm} 
\includegraphics[width=0.550\textwidth]{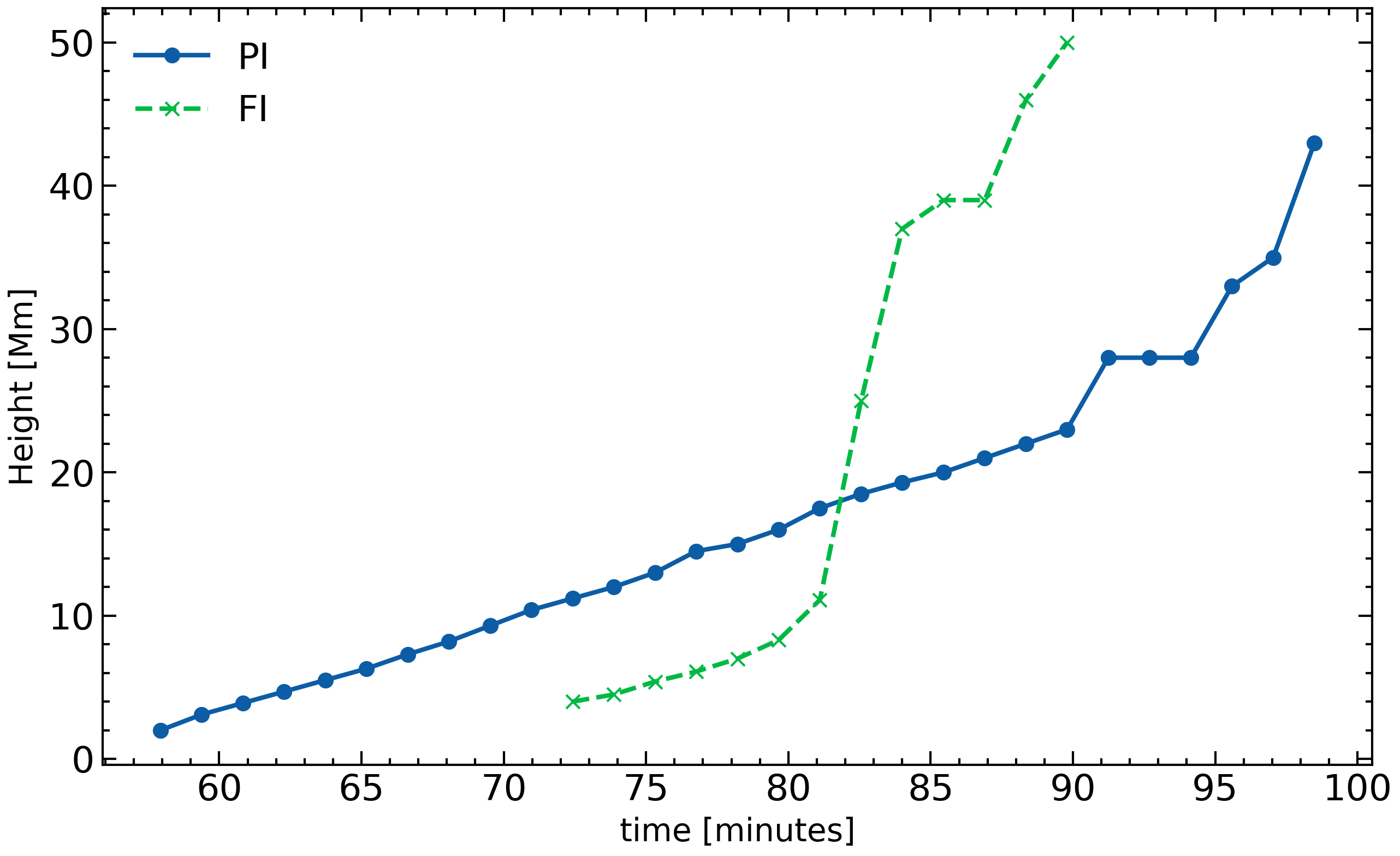}
\caption{Height time profile of the internal rising plasma in both simulations. The slow rise phase is prolonged in the PI case.}
\label{fig:fr}
\end{figure}

 The time evolution of the erupting magnetic field along height is shown in Figure \ref{fig:fr}. In fact, we follow the part of the rising plasma, at the vertical xz-midplane, which has the following properties: a) the azimuthal magnetic field component (Bx) is zero at that height and above and below it, Bx changes its sign, b) the magnetic field is directed almost solely along the y-direction and c) the fieldlines that are passing through this height have their photospheric footpoints passing through the maximum values (positive and negative) of Bz at the two main polarities of the emerging field. It seems that the plasma element which has these properties is located very close to the central part (the core) of the erupting field. We  find that in both cases, there is a slow rise phase and a fast rise phase, where the height increases exponentially over time. This height-time profiles indicate strongly that the erupting field in both cases erupt towards the outer solar atmosphere. There are two major differences in the two cases. In the PI case, the slow rise phase lasts longer (up to around $t = 90$ min) and the fast rise phase starts at a larger height, at around $z = 22$ Mm instead of $z \approx 6$ Mm for the FI case. To connect the height-time profile with the dynamics of the motion of 
the erupting field, we study the velocity profiles of the rising magnetized plasma in the next subsection. 

\vspace{1cm}
\subsection{The rising motion of the erupting field}
\begin{figure*}[p]
\centering
\includegraphics[width=.820\textwidth]{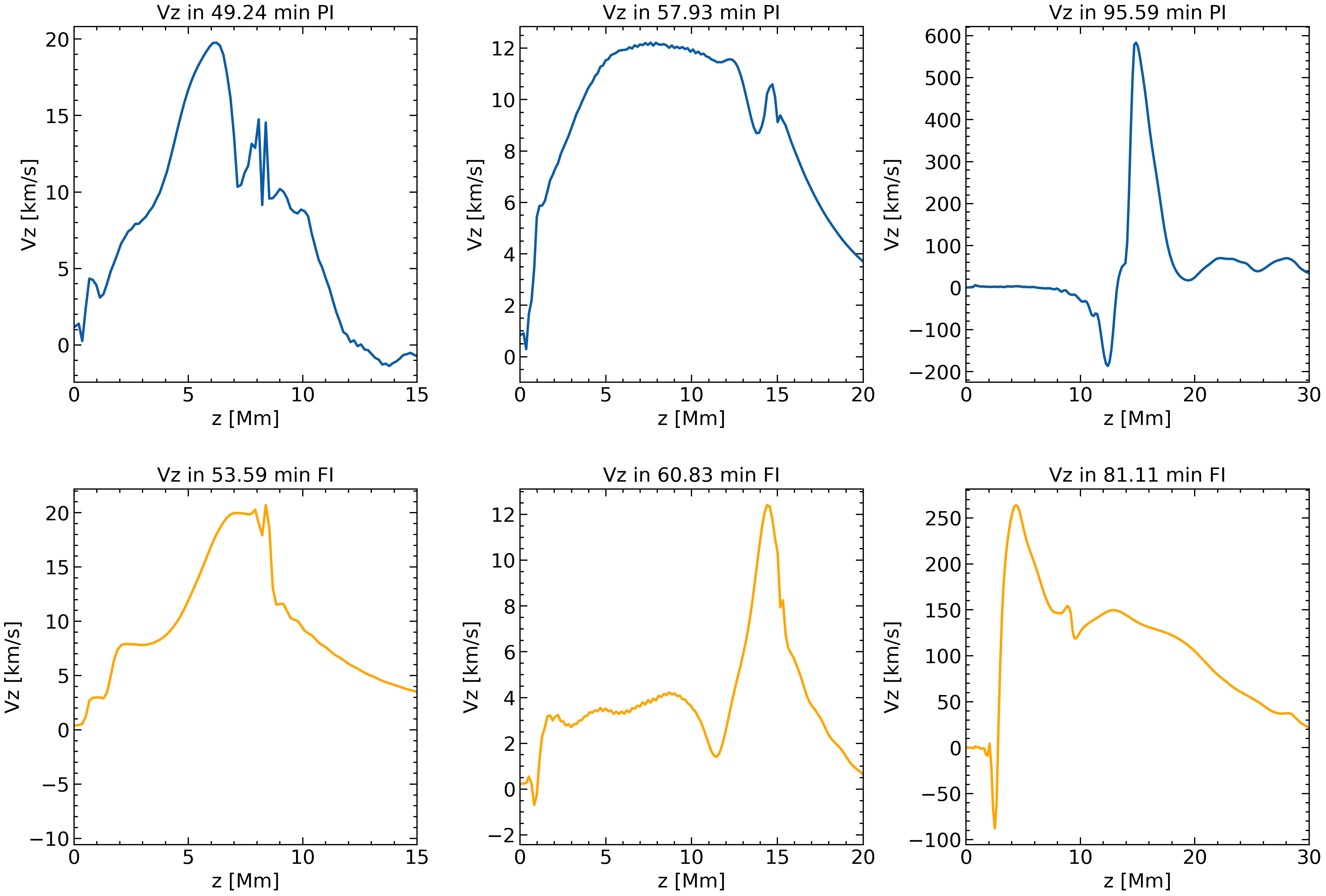}
\caption{The $\nu_z$ distribution in the center of the box along height for both simulations that depict the rising motion of the plasma  before the solar eruption (first and second collumns) and at the eruptive phase (third collumn).}
\label{fig:vz_slices}
\end{figure*}

Figure \ref{fig:vz_slices} shows the distribution of $\nu_z$ along height, at the center of the computational domain above the photosphere. We have chosen three times to show this distribution. First and second collumn shows $\nu_z$ before the eruption and the third collumn during the eruptive phase in both cases. 

In the FI case, at $t=53.59$ minutes the apex is located at $z \approx 8-9$ Mm. This is the early stage of emergence of the field above the photosphere and the plasma at the apex and inside the emerging field is moving with similar speed $\nu_z \approx 20 \, \text{km}\, \text{s}^{-1}$. At $t=60.83$ min, the emerging field expands well into the corona and the apex reaches $z \approx 14-15$ Mm moving with higher speed than the heavier plasma behind it. At $t=81.1$ min, the apex has reached the height of $z \approx 30$ Mm but now we find that the plasma inside the emerging field is moving faster than the apex. More precisely, we find that at the height of $z \approx 9-10$ Mm there is a local maximum in $\nu_z$, which means that the plasma is moving faster than any other plasma element above it. This fast moving plasma is the central part (at and around the core) of the erupting field. Underneath the erupting field, there is a strong bi-directional $\nu_z$ flow. This is because reconnection occurs there, which leads to the formation of the FR, which eventually erupts.

In the PI case, the rising motion of the erupting field shows some differences. At the early stage of emergence, at $t=49.24$ min, the plasma that is located between $z  \approx 4.5-7$ Mm is moving faster than the apex ($z \approx 8-9$ Mm). The faster moving plasma is actually the plasma, inside the emerging expanding volume, which is less heavy compared to the apex and, thus, it rises with higher speed. As time goes on, the apex reaches $z \approx 14-15$ Mm, and the plasma behind is still rising with higher speed. Notice that at this time, the amount of plasma which is rising with higher 
speed compared to the apex is larger, it extends from $z \approx 4.5-13$ Mm. This is because the arch-like structure of the magnetic field continues to stretch upwards and, thus, most of the plasma behind the apex moves faster. Also the fact that the plasma behind the apex is moving faster in the PI case is because the magnetic field emerges with an arch-like shape and the heavier plasma drains more effectively along the curved fieldlines due to gravity. In the FI case, the concave dips of the twisted fieldlines of the FR can carry more heavy plasma upwards. Here, it is worthwhile mentioning, that in Figure \ref{fig:pre1}  we had found heating of the plasma behind the apex of the emerging field. The reason for this heating is because the plasma is compressed at that interface, exactly between the faster moving plasma of the interior and the slower moving plasma of the apex. Still, at this stage there is no reconnection at the lower parts of the arch-like field. However, at $t=95.59$ reconnection occurs at $z \approx 11.5$ Mm and a bi-directional reconnection flow is created, with a maximum upward speed of $\nu_z \approx 580 \, \text{km}\, \text{s}^{-1}$. At this stage the emerging field has an overall flux rope-like configuration, which undergoes an eruptive phase. 
\vspace{1cm}
\subsection{The first eruption}
\begin{figure*}[p]
\centering
\includegraphics[width=1.1\textwidth]{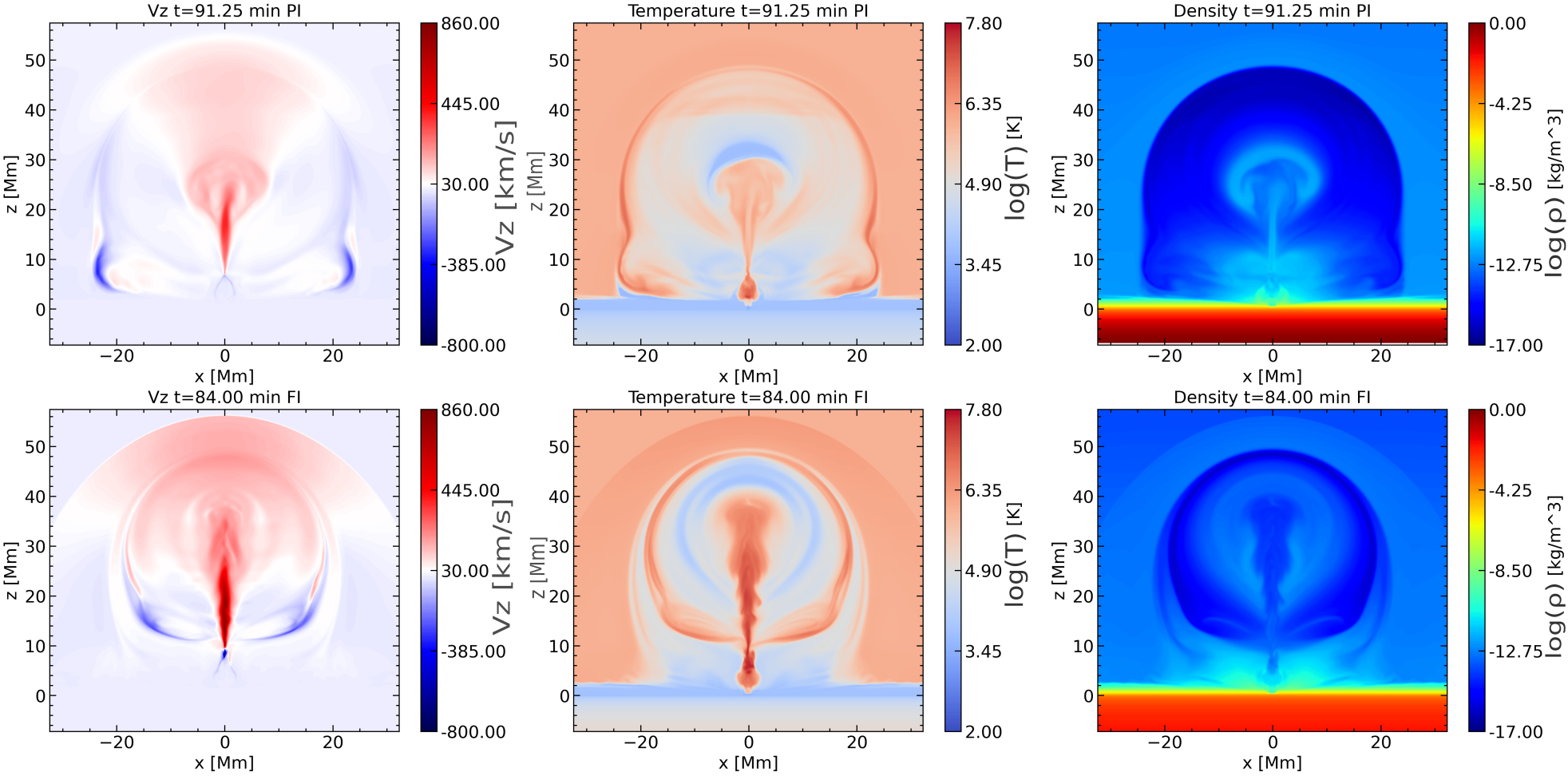}
\caption{Contour plots that depict the $\nu_z$ profile (first column), the logarithm of temperature (second column), and the logarithm of density (third column) in both simulations at the xz midplane. The contour plots are at times that capture the onset of the first solar eruption.}
\label{fig:firster}
\end{figure*}

To study the first eruption in the FI and PI experiments, we visualize the temperature, density and vertical velocity when the ``central" part of the erupting field is within the fast eruptive phase and the apex of the emerging field in the two experiments has reached similar heights. This is shown in Figure \ref{fig:firster}. The vertical velocity (Figure \ref{fig:firster} first column) show a strong upflow at the center of the domain. It extends from $z \approx 6-8$ Mm to the high corona and up to the very close proximity of the ``central" part of the erupting flux system. Basically, this is one part of the bi-directional reconnection flow, which occurs at the current sheet underneath the erupting magnetic field. The second part is a downflow (light blue color), which hits the heavier plasma at lower atmospheric heights. The downflows at the low edges of the expanding field (e.g. at $x=-20$ Mm, $x=20$ Mm and $z=4-6$ Mm for PI) is due to gravitational draining of the plasma, along the fieldlines, from their apex towards their footpoints.

The temperature distribution shows that the reconnection outflows are hot (as expected) and, thus, the plasma inside the expanding volume of the emerging field is a mixture of cool and hot elements within a temperature range of $10^4 - 10^7 K$. The density distribution (third collumn) correlates very well with the temperature distribution in both cases. By looking only at the density distribution, we
find that the central part of the erupting field in the PI case, at this time of the evolution and in coronal heights,  consists of more dense plasma. However,
over the whole time evolution of the magnetic flux system in our simulation, the average amount of density that is emerging or ejected (during the eruptions)
in the corona is higher in the FI case. Also the differences in the vertical velocity (e.g higher in the FI case) and temperature are due to the fact that at this stage of the evolution, the FI eruption is well into the fast rise phase (Figure \ref{fig:fr}), while the PI eruption is at the beginning of this phase.

In Figure \ref{fig:firster}, the central region of the emerged magnetic field dome may show evidence of the Kelvin-Helmholtz instability (KHI) in PI. This is particularly apparent in the central part or at the edges where velocity shear is present (first column, first plot in \ref{fig:firster}). The behavior of this instability in partially ionized plasmas has been extensively studied in previous works \citep{Ballester2018A&A...609A...6B, Zhao2018RAA....18...45Z, Li2018NatSR...8.8136L, Hilier2019PhPl...26h2902H, Yuan2019ApJ...884L..51Y}.
 We would like to highlight that this paper represents the first 3D study of solar eruptions with partially ionized plasma in a magnetic flux emergence experiment. Our primary focus is on the fundamental aspects of the dynamic evolution of these eruptions. While the study of KHI in partially ionized plasmas is indeed an interesting and important topic, it is outside the scope of our current work, which is dedicated to understanding the core dynamics of solar eruptions under the influence of partial ionization.
 \vspace{1cm}
\subsection{Energy evolution during eruptions}
\begin{figure*}[!ht]
\centering
\includegraphics[width=\textwidth]{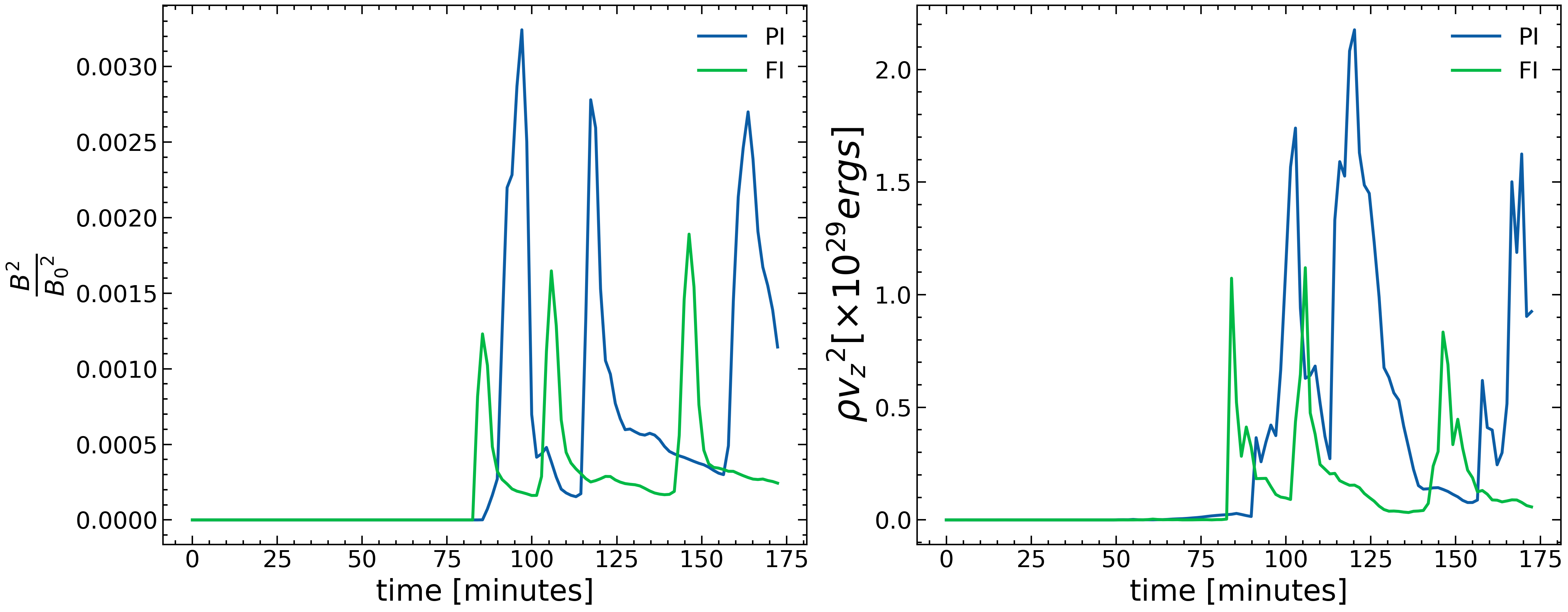}
\caption{Comparison of the average value of normalized $B^2$ at 40Mm height (left panel) and kinetic energy at the same height (right panel) in both simulations.}
\label{fig:kinetic}
\end{figure*}

Figure \ref{fig:kinetic} consists of two panels. The first panel shows the time evolution of the average value of $B^2$ at z=40 Mm at the xy plane, divided by the average value of $B^2$ at the subphotospheric tube at t=0. The second panel shows the temporal evolution of the kinetic energy integrated over a selected volume defined by a horizontal area of \(64.8\,\mathrm{Mm} \times 64.8\,\mathrm{Mm}\) and a vertical thickness of \(0.154\,\mathrm{Mm}\) centered at a height of \(40\,\mathrm{Mm}\). Formally, the kinetic energy \(E_{\mathrm{kin}}\) at time \(t\) is given by:
\[
E_{\mathrm{kin}}(t) = \int\!\int\!\int \tfrac{1}{2}\,\rho\,v_z^2 \,dx\,dy\,dz,
\]
where the integration covers the specified volume.
 The right panel shows three local maxima in PI and three local maxima in FI. The first maximum in FI occurs between \(t=80\) and \(t=90\) minutes, corresponding to the reconnection outflow that triggers the first eruption. In contrast, this event occurs in PI between \(t=100\) and \(t=110\) minutes. The second maximum in FI, occurring between \(t=100\) and \(t=110\) minutes, corresponds to the reconnection outflow initiating the second eruption. The third maximum in FI, marking the onset of the third eruption, occurs between \(t=140\) and \(t=150\) minutes. 

\begin{figure*}[!ht]
\centering
\includegraphics[width=\textwidth]{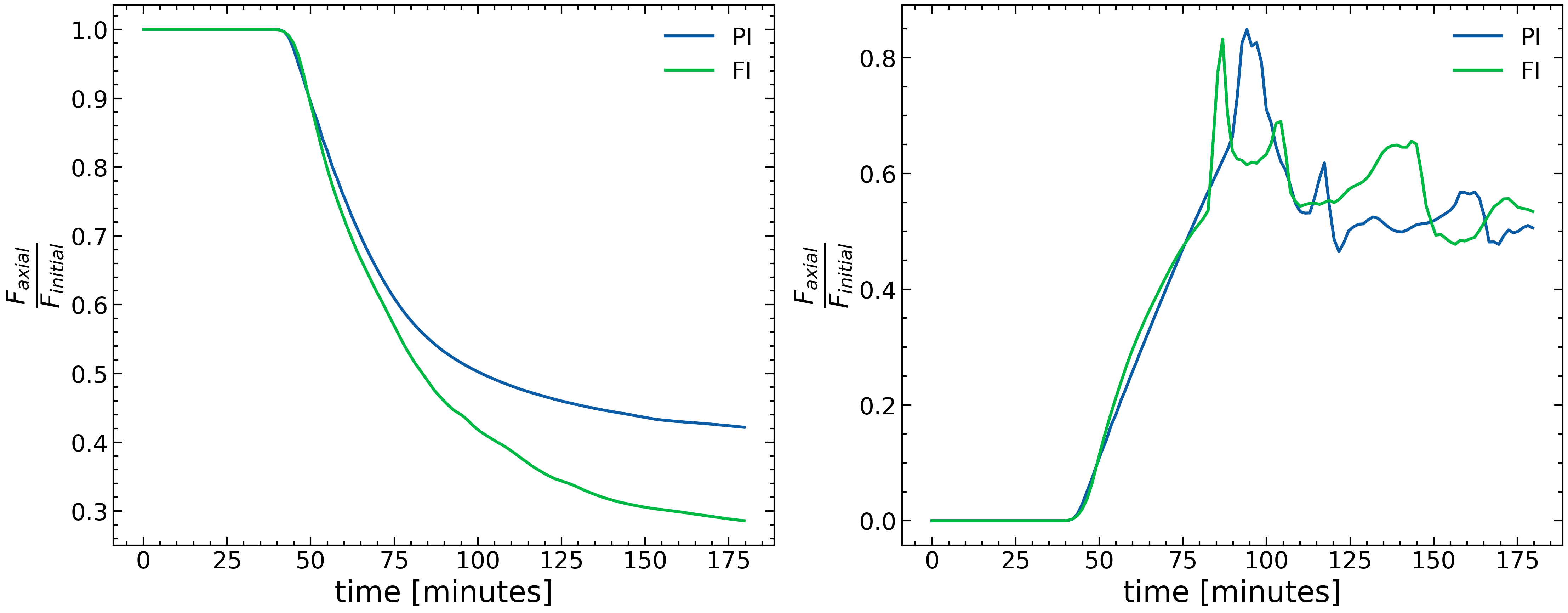}
\caption{Comparison of the normalized axial flux below photosphere (left panel) and above photosphere (right panel) in both simulations.}
\label{fig:axial_flux}
\end{figure*}

Meanwhile, in PI, the second maximum occurs between \(t=115\) and \(t=130\) minutes and corresponds to the reconnection outflow that signifies the onset of the second eruption, which appears to be the strongest. The third maximum in PI, occurring between \(t=160\) and \(t=180\) minutes, corresponds to the third eruption.
 As we have explained in the previous sections, there is a small time difference on the time of occurence of emergence and eruptions between the PI and FI case. The actual emergence to the corona proceeds a bit earlier in the PI case, since the formation of the new FR that eventually erupts in the corona, it occurs earlier in the FI case. By looking at the left panel in Figure \ref{fig:kinetic} we find that magnetic energy reaches three local maxima at times just before the maxima of the kinetic energy. This is because on every eruption, the strong magnetic field of the erupting FR reaches first the corona, followed by the strong vertical reconnection upflow, which originates at the vertical current sheet behind the 
FR. Thus, there is a small delay between the maxima of magnetic energy and the maxima of the kinetic energy. Another intersting result is that the (normalized) magnetic energy is higher in the FI case as a whole. Therefore, the FI eruptions consist of stronger magnetized plasma. 
In turn, this is because (see Paper I) stronger magnetic field, including the axis of the emerging flux tube, rises above the photosphere in the FI case, while in the PI case the strong axial magnetic field does not succeed to emerge above the photosphere. Now, by looking at the right panel in Figure \ref{fig:kinetic} we find that the plasma acceleration during the eruptions in the FI case is less compared to the PI case. This is due to the larger value of $\nu_z$ in the FI case. The main component of the vertical velocity $\nu_z$ is the reconnection upflow, which follows the core of the erupting FR. The reconnection, which leads to the formation of the erupting FRs, occurs at different heights in the two cases. For the FI case, it occurs in the low atmosphere (chromosphere/ transition region), while for the PI cases it happens in the corona. Because the reconnection outflows move at speeds close to the local Alfvén speed, and the latter is higher in the corona, the kinetic energy of the eruptions in the PI case is higher.    

\begin{figure*}[htbp]
    \centering
    \includegraphics[width=1.090\textwidth]{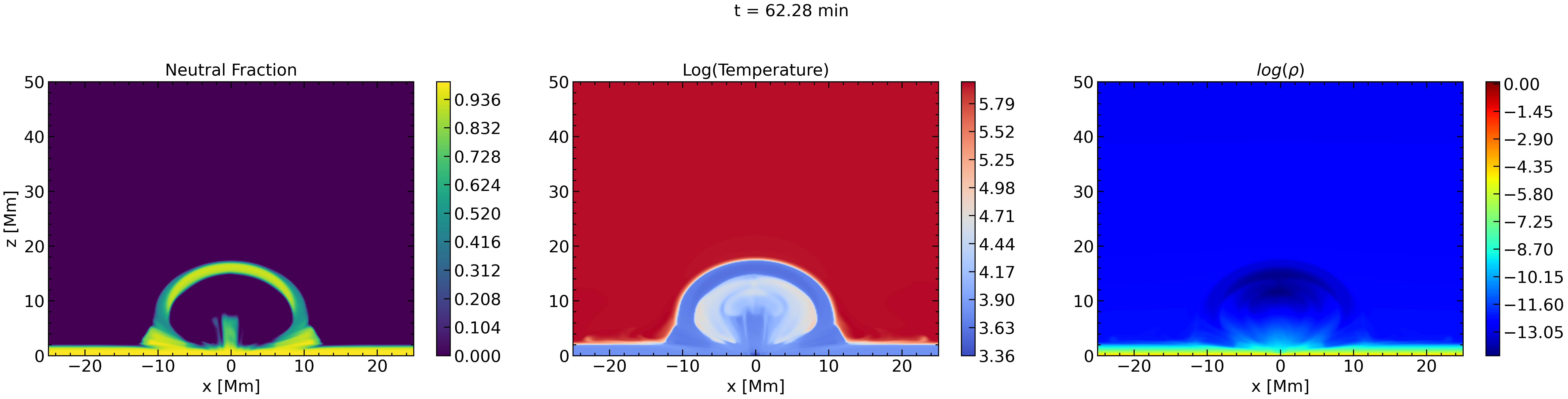}
    \hfill

    \includegraphics[width=1.090\textwidth]{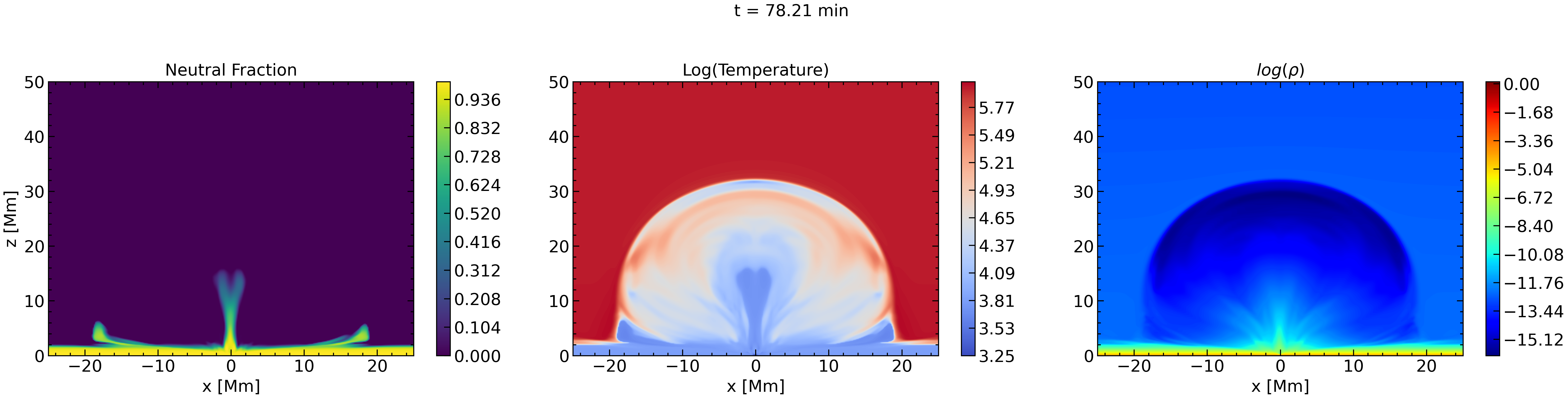}
    \hfill

    \includegraphics[width=1.090\textwidth]{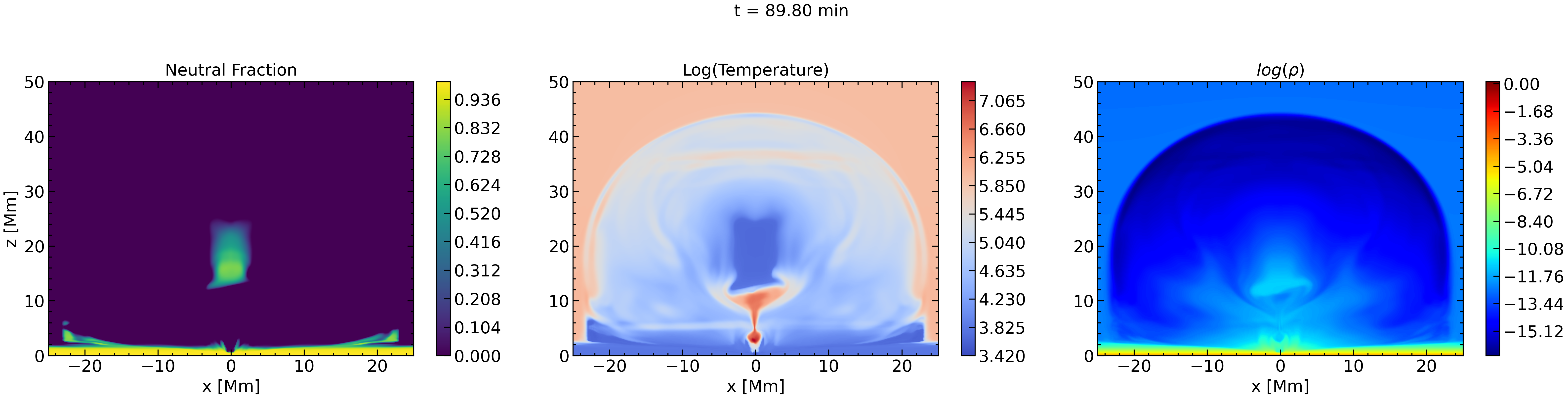}
    
    \caption{Contour plots of the neutral fraction, the logarithm of the temperature and the logarithm of the density in the xz mid plane at t=62.28 minutes, t=78.21 minutes, and t=89.80 minutes.}
    \label{fig:neuts}
\end{figure*}
We now focus on how much axial flux stays below the photosphere and how much axial flux is found in the corona during the simulation. More precisely, we calculate $F_{\text{axial}} = \iint By \, dx \, dz$ at the vertical xz-midplane at y=0. These values were normalized with the initial flux at \(t=0\) for two distinct regions: below the photosphere (left panel in Figure \ref{fig:axial_flux}) and above the photosphere (right panel). In the PI simulation, we observed a significant portion of the axial flux remains subphotospheric. This observation aligns with the findings from Paper I, where the axis of the flux tube fails to emerge fully at the solar surface. 
We find that, towards the end of the simulation, the normalized axial flux in the PI case converges to a value of about $40\%$ and the FI case to a value just below $30\%$. Above the photosphere (right panel in Figure \ref{fig:axial_flux}), we find a similar evolution in both cases. The flux reaches three local maxima, in the PI and in the FI simulations, at times just before the eruptions. This indicates that the structures that erupt have a considerable axial field (By) component. This is in agreement with the results presented in the previous section (3.2) where 
we have traced the fieldlines before and after the first eruption of the magnetic field. There, we have shown that the erupting field adopts the configuration of a FR with magnetic fieldlines directed along the y-axis at the core of the FR.
Therefore, the local increase of the normalized axial flux is related to the formation of new FRs, which eventually erupt into the outer solar atmosphere. 
Overall, we find that the normalized axial flux above the photosphere is higher in the first eruption and it gradually decreases over the 
simulation. This might indicate that the following eruptions do not have the same exact structure with the first eruption. 
Another interesting result is that the values of the normalized axial flux vary over a similar range during the simulation. For instance, the first maxima is close to the value of 0.85 and it seems to approach a saturation stage towards the end of the simulation, with a value of about 0.5. Thus, in both experiments, the amount of normalized axial flux found in the corona does not differ dramatically over the whole time of the simulation. A more detailed investigation is required to study both, the structure of the erupting field (e.g. \citet{Moreno-Insertis_etal2013}) and the evolution of the associated axial flux.
\vspace{1cm}
\subsection{Neutrals, emergence and eruptions}
In this section, we study how the neutral fraction evolves during the emergence and the first eruption in the PI case. Figure \ref{fig:neuts} consists of three columns, neutral fraction, temperature and density (from left to right) at the vertical xz-midplane of the numerical domain at four different times.

In the first row, the temperature distribution shows that there are three interesting areas within the emerging volume. The top arch-like layer with cool plasma, the middle/central column where the plasma is also cool and the interface layer between them, where the plasma has much higher temperature. Basically, this stage of evolution has been discussed in Figure \ref{fig:pre1}, where we had pointed out that the less dense rising plasma inside the expanding volume of the field moves faster and it compresses the plasma above it, causing heating at the interface. Now, we
find that the neutral fraction is higher at both areas with low temperature: at the arch-like front and at a vertical column at the center (x=0,y=0) of the domain. It also extends in the y-direction along the PIL of the emerging bipolar field. We should mention that this column of neutrals is between the footpoints of the J-like fieldlines, like the red fieldines, top row Figure \ref{fig:morph}, which will eventually reconnect to form the erupting FR. 

As the time goes on (second and third row in Figure \ref{fig:neuts}), the whole emerging volume expands laterally, but also, the faster moving plasma continues to move upward, it compress further the material above it, and thus it's temperature increases. As a result of the higher temperature at the top arch-like layer, neutrals continue to exist only at the vertical middle column, where the plasma is still cool and dense. Notice that, the lower part of that column becomes thinner. This happens because the oppositely directed magnetic field (J-like fieldlines) in the close vicinity of the column, on the left side (negative x-axis) and on the right side (positive x-axis), get progressively closer together and a strong and thin current sheet is build up, where the local temperature increases. During this procees, low-atmospheric dense plasma (at around x=0, y=0) is brought up, loading part of the neutrals column with heavy material.

At t=89.80 min (third row) the reconnection between the J-like fieldlines has already started. The temperature distribution shows the cool area of the internal magnetic field and the hot reconnection plasma flows underneath it. The density correlates very well with the temperature distribution. Neutrals now exist only within the cool area of the internal magnetic field, which erupts. A considerable amount of neutrals is carried upwards, well inside corona, and just above the hot jets. The largest fraction of the neutrals is also very
dense. Eventually (not shown here) and as the eruptive plasma rises higher, neutrals continue to exist although to a lower fraction, as they are pushed upwards by the upward reconnection jet. Neutrals disappear when the temperature within the eruptive material increases enough to prohibit their existence.
\vspace{1cm}
\subsection{Discussion on the Equation of State}

We are using the same equation of state (EOS) as \citet{Leake_etal2013a}. In fact, our work is the first 3D study using this EOS. Briefly, we include the effects of ion-neutral collisions on the magnetic field evolution in the induction equation by adding the ambipolar diffusion term. We also account for the ionization-recombination of neutral atoms in the EOS by adding the second term on the right-hand side of the internal energy equation (Eq. 9 in Paper I). The addition of these terms includes the effects of partially ionized plasma in our PI experiment and differentiates it from FI. Our simulations are single-fluid MHD with a hydrogen atom EOS. 

This specific EOS should be used with caution, as studies using an EOS that includes multiple different atoms and electron donor atoms \citealp{Bifrost_2020A&A...638A..79N} showed that a ``simple'' EOS with hydrogen could potentially overestimate the ambipolar diffusion coefficient, particularly in regions with low temperature and low density (see Figure 2 in \citet{Nobrega2020A&A...633A..66N} and \citealp{Rempel2021ApJ...923...79R}). This overestimation could lead to large ion-neutral drift velocities ($V_d$), which could potentially create heating that might be misinterpreted. We highlighted in Paper I that the heating from ion-neutral collisions was not powerful enough to heat the local plasma above nominal values but was sufficient to mitigate the cooling from the adiabatic expansion of the magnetic loop in the solar atmosphere. Additionally, we found that ambipolar diffusion had no effect on the emerged vertical magnetic flux (Paper I). Similar results have been found by \citet{Nobrega2020A&A...633A..66N} using a more sophisticated EOS.

More recent work by \citet{Hillier2024RSPTA.38230229H} has edited the analytic formulation of the ambipolar diffusion approximation to accommodate large $V_d$ (e.g., 40\,km\,s$^{-1}$, as observed by \citet{Nobrega2020A&A...633A..66N} in regions with nearly 2000\,K). They managed to decrease these $V_d$ values to 20\,km\,s$^{-1}$.

\begin{figure}[!ht]
\centering
\hspace*{-1.5cm} 
\includegraphics[width=0.550\textwidth]{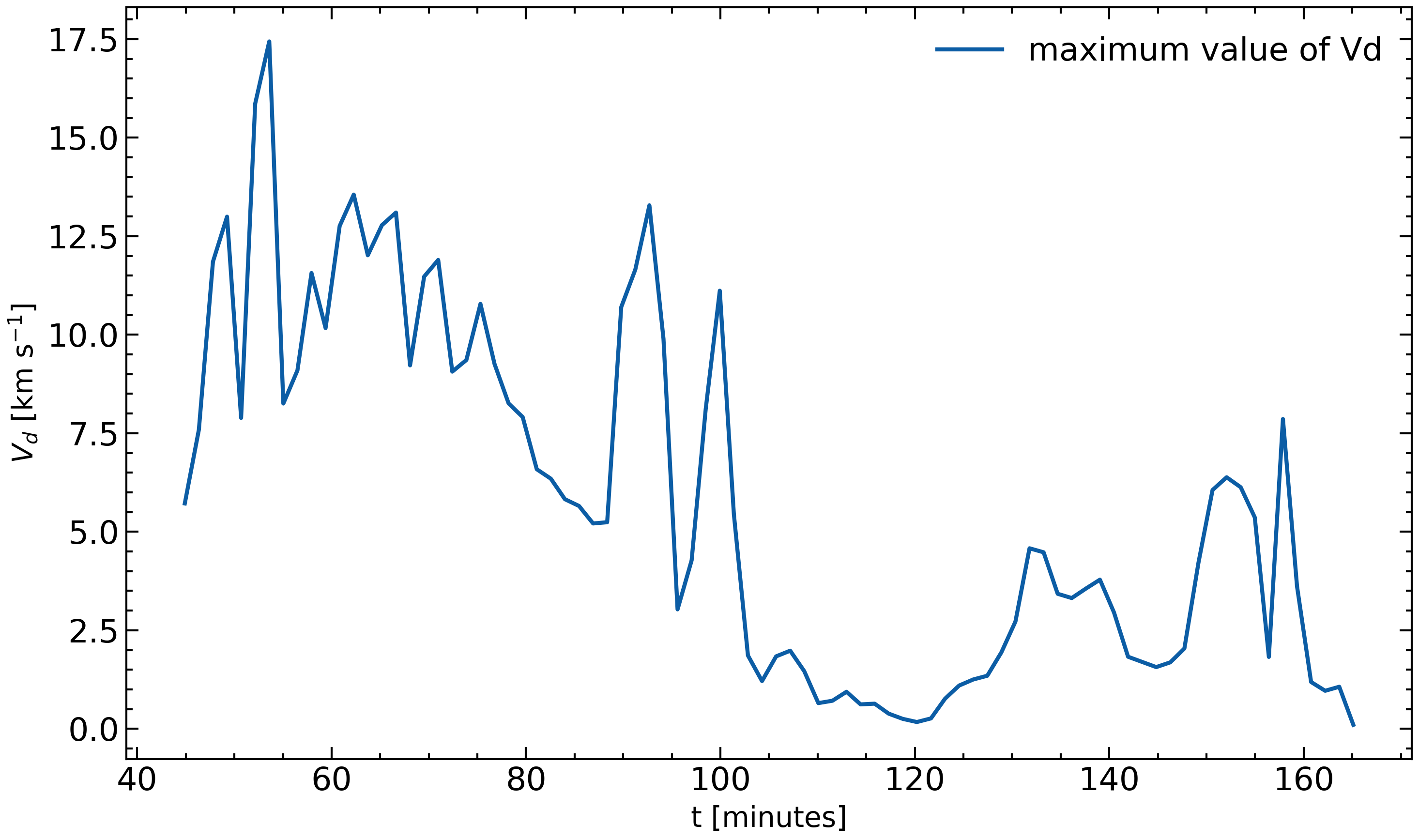}
\caption{Temporal evolution of the maximum value of $V_d$ above the photosphere for the PI simulation.}
\label{fig:vd}
\end{figure}

Using the formula for $V_d$ from \citet{Nobrega2020A&A...633A..66N}:
\begin{equation}
\mathbf{V}_d = \eta_\perp \frac{\mathbf{J} \times \mathbf{B}}{\mathbf{|B^2|}},
\end{equation}
we plotted in Figure \ref{fig:vd} the temporal evolution of the maximum value of $V_d$ above the photosphere for the entire simulation. 

The range of values indicates that throughout our simulation, $V_d$ is less than 17.5\,km\,s$^{-1}$, which is classified as a low $V_d$ range. Thus, the standard ambipolar diffusion approximation is sufficient. 

\vspace{1cm}
\section{Conclusions}
We have performed 3D MHD simulations, including partial ionization, to study its effect on the process of magnetic flux emergence above the photosphere.  Our initial study (Paper I) focused on how partial ionization affected the tube's emergence beneath the solar surface. The present study elaborates the implications of partial ionization on the structure, dynamics and thermodynamical aspects of the emergence into the solar corona, and on the eruptions that 
originate on the emerging bipolar region.
\par In our previous study (Paper I), we showed that the axis of the tube remains below the photosphere in the PI simulation. Here, we show that the emerging field above the photosphere consists of arch-like fieldlines, with no or very little twist. After the expansion into the outer solar atmosphere, reconnection between J-like fieldllines occur at the central part of the domain and at coronal heights. Due to this reconnection, a new FR is formed, which eventually erupts in an ejective manner. Therefore, the topology of the fieldlines, the overall structure of the emerging field and the process leading to the formation of an erupting FR is distinctly different between the PI and the FI cases. We should mention, that in the FI cases, the axis of the original twisted flux tube emerges above the photosphere and stays there for the rest of the simulation. Sheared fieldlines, which belong to this tube, experience reconnection at the PIL of the emerging region, which leads to the formation of a new FR that erupts into the corona.

From the work by \citep{Wragg1981SoPh...70..293W,Reale2010LRSP....7....5R}, our emerged magnetic domes exhibit lengths, densities, and temperatures similar to the active region loops presented in the referenced works. Thus, these emerged domes in our simulations align with the observations of active region loops.
Figure \ref{fig:vz_slices} depicts the vertical velocity at the center of the box at three distinct times. The first column shows the vertical velocity of plasma elements along the vertical line located at the center of the box. The plasma element at the apex of the tube in the PI scenario has a vertical velocity of approximately $\nu_z \approx 15 \, \text{km}\, \text{s}^{-1}$. In the FI scenario, the same plasma element at the apex has $\nu_z \approx 20 \, \text{km}\, \text{s}^{-1}$. These values are similar to those reported by \citet{Nobrega-Siverio_etal2016}
\par We have confirmed the results by \citep{Leake_etal2006,Arber_etal2007,Martinez-Sykora_etal2012,Leake_etal2013b,Nobrega2020A&A...633A..66N} that the plasma inside the expanding magnetized volume in the PI simulations is less cool than in the FI simulation due to ion-neutral collisions, which mitigate the intensive cooling due to the adiabatic expansion (PI simulation). Also, the plasma inside the expanding volume and under the apex of the magnetic field is less dense, compared to the FI simulation, and it rises with higher speed than the apex. As a consequence, we find that the plasma just below the apex is compressed and its temperature increases. Therefore, the temperature and density distribution inside the expanding magnetic field is different in the two cases.
\par During the time of our simulations, we find that there are three major eruptions following the expansion of the emerging field. The normalized 
axial flux, which remains below the photosphere is less in the PI case, which is in agreement with the afore-mentioned results. However, the amount of 
normalized axial flux above the photosphere is similar in both simulations. Further, more detailed analysis is required to study the nature of this result, which is presumambly related to the formation mechanism of the eruptions and how much flux they carry during their formation and ejection. We should mention that we have calculated the average values of the normalized $B^2$ and $\rho \nu_z{^2}$ at the high corona. We have found that there are local maxima of these quantities during the eruptions. Overall, in the PI case, the local maxima are lower for $B^2$ and higher for $\rho \nu_z{^2}$. The later indicates that the PI eruptions consist of weaker magnetized plasma (relative to the initial magnetic field of the sub-photospheric tube) and they are faster by a factor of two.
In our computation of the normalized axial flux, we align our findings with those of \citet{Moreno-Insertis_etal2013}, who also calculated the normalized axial flux. In their Figure 14, it is shown that approximately 30\% of the axial flux remains below the photosphere, which is consistent with our results in the FI simulation. However, in the PI simulation, we observe a different outcome: around 45\% of the axial flux remains below the photosphere due to the axis staying subphotospheric.

\par For the first eruption in the PI case, we have found that there is a thin column of neutrals, at the central region of the domain, extending from the PIL up to the 
low corona. There the plasma is cool and relatively dense. Eventually, a strong current sheet is build up in this area and reconnection of J-like fieldlines occurs to form the first FR that erupts towards the outer solar atmosphere. The reconnection upflow at the current sheet is pushing upwards a fraction of the neutrals together with the newly formed FR. Eventually, the hot reconnection flows increase the plasma temperature close to the erupting FR, which in turn diminishes the amount of neutrals in the high atmosphere. 

In our simulations, the emergence occured into a null corona. It is well known that the emergence into a magnetized atmosphere could lead to the formation of solar jets, which could be triggered by reconnection or driven by eruptions (e.g. \citet{Raouafi_2016SSRv..201....1R}). In a forthcoming study, we will examine the effect of partial ionization in the structure and thermo-dynamical aspects of jets in the solar atmosphere. 

\section*{Acknowledgements} 
The authors acknowledge support from the Royal Society grant RGF/EA/180232. This research has been supported by the European Research Council through the Synergy grant No. 810218 (“The Whole
Sun,” ERC-2018-SyG);  The work was supported by the High Performance Computing facilities of the University of St. Andrews “Kennedy”.

\bibliographystyle{apj}
\bibliography{bibliography}

\clearpage

\end{document}